Article

# Fused inverse-normal method for integrated differential expression analysis of RNA-seq data

Birbal Prasad[1] and Xinzhong Li[1*]

[1]National Horizons Centre, School of Health and Life Sciences, Teesside University, Darlington, DL1 1HG, UK. Email: B.Prasad@tees.ac.uk; X.Li@tees.ac.uk
*Correspondence: Email: X.Li@tees.ac.uk, Phone: +44-01642738451

## Abstract

**Background:** Use of next-generation sequencing technologies to transcriptomics (RNA-seq) for gene expression profiling has found widespread application in studying different biological conditions including cancers. However, RNA-seq experiments are still small sample size experiments due to the cost. Recently, an increased focus has been on meta-analysis methods for integrated differential expression analysis for exploration of potential biomarkers. In this study, we propose a p-value combination method for meta-analysis of multiple related RNA-seq studies that accounts for sample size of a study and direction of expression of genes in individual studies.

**Results:** The proposed method generalizes the inverse-normal method without increase in computational complexity and does not pre- or post-hoc filter genes that have conflicting direction of expression in different studies. Thus, the proposed method, as compared to the inverse-normal, has better potential for the discovery of differentially expressed genes (DEGs) with potentially conflicting differential signals from multiple studies related to disease. We demonstrated the use of the proposed method in detection of biologically relevant DEGs in glioblastoma (GBM), the most aggressive brain cancer. Our approach notably enabled the identification of over-expression in GBM compared to healthy controls of the oncogene *RAD51*, which has recently been shown to be a target for inhibition to enhance radiosensitivity of GBM cells during treatment. Pathway analysis identified multiple aberrant GBM related pathways as well as novel regulators such as *TCF7L2* and *MAPT* as important upstream regulators in GBM.

**Conclusions:** The proposed meta-analysis method generalizes the existing inverse-normal method by providing a way to establish differential expression status for genes with conflicting direction of expression in individual RNA-seq studies. Hence, leading to further exploration of them as potential biomarkers for the disease.

## Background

RNA sequencing (RNA-seq) technologies are now increasingly considered for whole transcriptome gene expression quantification studies as compared to traditional microarray technologies due to its high technical reproducibility and greater resolution [1]. Over the last decade, it has found widespread application in studying different biological conditions including cancers. For instance, sequencing data archived on The Cancer Genome Atlas (https://portal.gdc.cancer.gov/) have been used in a number of studies to explore potential biomarkers and mechanisms in oncogenesis [2][3]. Despite its advantages and few large RNA-seq datasets [4][5], RNA-seq experiments are still small sample size experiments because of its high cost. This leads to a problem of reduced statistical power in studies such as differential expression analysis where thousands of genes are studied at a time but only have tens to hundreds of samples. Combination of data or results from multiple independent but related studies (referred to as meta-analysis) have been widely used to increase available sample size and consequently the statistical power to obtain a precise estimate of gene expression differentials [6][7]. In the context of differential expression analysis, several different meta-analysis approaches have been proposed for integrating microarray studies [8][9] and some of them have later been adapted for RNA-seq data [10][11].

For microarray gene expression studies, apart from vote-counting and direct merging of datasets, meta-analysis methods can mainly be classified into three types based on the combined statistic [7]. First are methods based on effect sizes combination in which a combined effect (for instance, strength of differential expression between two conditions for a gene) is obtained based on the calculated effect sizes and its variance where two possible models namely, fixed and random effects model are used to obtain the combined effect [12]. Second are approaches based on integration of p-values obtained from per-study analysis into a single combined p-value per gene [13]. Lastly, are approaches based on rank combination which are non-parametric and allow for integration of studies based on a statistic that can be ordered, e.g., fold change of a gene [14]. However, RNA-seq data are counts data, i.e., normalized number of sequenced reads within a certain gene or transcript, unlike the microarray data which are continuous, e.g., normalized signal intensity of image [15]. Hence, the methods initially proposed for microarray data are not suited to be applied directly to RNA-seq data in many cases [10].

In case of RNA-seq data, Poisson or Negative-Binomial distributions are typically used to model gene counts [16]. Kulinskaya et al. (2008) [17] described an effect-size combination method using an Anscombe transformation of Poisson distributed data. However, as highlighted by Rau et al. (2014) [10], this effect-size combination approach is not appropriate for RNA-seq data due to over-dispersion among biological replicates and presence of zero-inflation. Rau et al. (2014) considered two p-value combination methods, namely Fisher and inverse normal (IN) or Stouffer's methods, previously proposed and used for meta-analysis of microarray studies [8][9][13] and demonstrated how these can be adapted in RNA-seq data analysis. Their results illustrated that Fisher and IN methods were very similar to each other in terms of performance but were better than the global and per-study differential analysis [10]. These two (Fisher and IN) p-value combination approaches have been implemented in several R packages, e.g., metaRNASeq [10], metaseqR [18] and metaSeq [19] and are most widely used methods due to its statistical simplicity and ease of direct application for meta-analysis of RNA-seq studies for differential expression.

Among all the existing meta-analysis methods for RNA-seq data discussed above, only few of the p-value combination methods (e.g., IN and PANDORA p-value [18]) allow for incorporation of information regarding the number of replicates in different studies to be combined through specification of a set of weights. However, information related to the direction of expression (up- or down-regulated) of a gene across different studies is not accounted for or included in any of these meta-analysis methods for RNA-seq data. Under- and over-expressed genes are analysed together and genes exhibiting conflicting direction of expression across studies are either removed prior to meta-analysis or are suggested to be identified and removed post-hoc [10][11]. Hence, no conclusion can be drawn with regards to differential expression for the genes that have conflicting direction of expression across different studies. Given that a significant proportion of genes may exhibit conflicting direction of expression across different gene expression studies [20], particularly when more and more RNA-seq data are publicly available and included into integration, emphasis is warranted on including this important prior information in a meta-analysis setting.

Recently, importance of inclusion of direction of expression information for genes in RNA-seq meta-analysis has been recognized leading to a generalization of some existing p-value combination methods such as Fisher method and Bayesian Hierarchical Model [21][22][23]. However, these generalizations come at a cost of increased statistical and computational complexity which discourages their widespread application to transcriptomic studies. In this study, we aimed to develop a new approach for integrated differential meta-analysis of RNA-seq data which accounts for both the sample size and direction of gene regulation in each study. The proposed approach leads to a generalization of the IN method without introducing additional cost and hence is simple and intuitive for real data application. First, we propose a modified inverse-normal (MIN) approach for p-value combination and assess its performance by comparing it with the IN method based on an extensive simulation study. Next, to overcome the conservative nature of MIN method, we further propose a fused inverse normal (FIN) method for p-value combination and assess its performance by comparing it to IN and MIN methods in a simulation study. Then an application to a set of real glioblastoma (GBM, the most aggressive type of brain cancer) studies has been conducted. Moreover, we assessed the relevance of the identified differentially expressed genes (DEGs) for GBM by using Ingenuity Pathway Analysis (IPA, www.qiagen.com/ingenuity) for pathway analysis and upstream regulator analysis.

## Methods

Let $y_{gcrs}$ be the observed count for gene $g$ $(g = 1, 2, \dots, G)$ in condition $c$ $(c = 1, 2)$ of biological replicate $r$ $(r = 1, 2, \dots, R_{cs})$ in study $s$ $(s = 1, 2, \dots, S)$. For an integrated differential analysis of gene expression across multiple studies, we first conducted the differential expression analysis within a given study $s$ using edgeR package (version 3.26.5) in R version 3.6.0 [24] with likelihood ratio test as the test for differential expression. Let $p_{gs}$ be the raw p-value for per-gene and per-study obtained using the individual differential expression analysis within a given study $s$ for gene $g$. The null hypothesis tested in the individual differential analysis is that the gene is non-differentially expressed in the particular study. For notational convenience, the notations similar to the ones used in Rau et al. (2014) [10] were adopted in this study.

### Modified inverse-normal method

Let $B_{gs}$ be a Bernoulli random variable which takes values 1 and -1 when a gene $g$ is over and under expressed respectively in a study $s$. A gene can be assessed as over- or under-expressed based on the fold change values (>1 or <1) of the gene in a study. Then, for a gene $g$, we define a combined statistic

$$N_g = \sum_{s=1}^{S} w_s B_{gs} |\Phi^{-1}(1 - p_{gs})| \tag{1}$$

where $w_s$ are a set of study specific weights described by Marot and Mayer [25] as follows:

$$w_s = \sqrt{\frac{\sum_c R_{cs}}{\sum_k \sum_c R_{ck}}} \tag{2}$$

Here, $\sum_c R_{cs}$ is the total number of biological replicates in a study $s$ for all condition $c$ and $\sum_k \sum_c R_{ck}$ indicates the total number of biological replicates in all studies. Moreover, $N_g$ can be considered as a weighted z-score. An advantage of this weighting criteria is that larger weights are attributed to studies with larger sample sizes. $\Phi$ is the standard normal cumulative distribution function and $p_{gs}$ is the raw p-value obtained for gene $g$ by differential analysis for study $s$.

It is assumed that $p_{gs}$ are uniformly distributed under the null hypothesis and $\Phi^{-1}(1-p_{gs})$ is standard normal in the above formula (1), but this assumption of $p_{gs}$ is not automatically satisfied when dealing with RNA-seq data [10]. However, filtering of very low expressed genes in each study results in p-values which are roughly uniformly distributed under the null hypothesis [10]. Then, we have that $B_{gs}\left|\Phi^{-1}\left(1-p_{gs}\right)\right| \sim N(0,1)$ (see Theorem 1).

**Theorem 1:** Let $X$ and $Y$ be two independent random variables where $X \sim N(0,1)$ and $Y$ is a Bernoulli random variable taking values 1 and $-1$. Then, $Z = Y|X|$ is standard normal distributed.

**Proof:** Using the first principle,

$$\begin{aligned}\mathbb{P}[Y|X| \leq t = \; & \mathbb{P}[Y = 1, |X| \leq t] + \mathbb{P}[Y = -1, -|X| \leq t] \\ = \; & \tfrac{1}{2}\mathbb{P}[|X| \leq t] + \tfrac{1}{2}\mathbb{P}[|X| \geq -t] \end{aligned} \qquad (3)$$

Now, if $t < 0$, the RHS of (3) becomes $\frac{1}{2}\mathbb{P}[|X| \geq t]$. By symmetry of the normal distribution, we have

$$\mathbb{P}[Y|X| \leq t] = \mathbb{P}[X < t] = \Phi(t)$$

where $\Phi$ is the cumulative distribution function of standard normal.

For $t \geq 0$, the RHS of (3) becomes $\frac{1}{2}\mathbb{P}[|X| \leq t] + \frac{1}{2}$. Hence, by symmetry of the normal distribution, we have

$$\mathbb{P}[Y|X| \leq t] = \mathbb{P}\big[X \in [0,t]\big] + \frac{1}{2} = \mathbb{P}[X \leq t] = \Phi(t)$$

Thus, $Z \sim N(0,1)$. ■

Hence, $N_g$ in equation (1) is a linear combination of independent standard normal variables. Thus, is also standard normal. A two-sided test can then be performed with $H_0$ being that the gene $g$ is not differentially expressed between two conditions (case vs control) and combined p-value is given by $(p_g = \mathbb{P}\big(|z| \geq N_g\big)$, i.e.

$$p_g = 2[1 - \Phi(|N_g|)]$$

A correction for multiple testing to control the false discovery rate (FDR) at a desired level $\alpha$ can be done by Benjamini-Hochberg (BH) approach [26].

**Simulation study: MIN and IN comparison**

To investigate the performance and compare the MIN method with state-of-the-art p-value combination method (IN with post-hoc filtering), we performed a simulation study. An extensive set of RNA-seq data was generated using the negative binomial distribution for the counts $y_{gcrs}$ and method described in Rau et al. (2014) [10] (see Supplementary 1, section: Simulation study model). Parameters for the simulation study were estimated from a real RNA-seq dataset for Alzheimer's disease (AD) study downloaded from Gene Expression Omnibus (GEO, https://www.ncbi.nlm.nih.gov/geo/) [27] with accession number GSE125583 which contains data for 219 AD patients and 70 normal control subjects. The method used for estimation of mean and dispersion parameters from GSE125583 were as described in Rau et al. (2014) with BH p-value < 0.05 being used to classify a gene as a DEG. This dataset was chosen as it has considerable number of samples for both biological conditions, namely case and control. Simulation settings for inter-study variability parameter ($\sigma$), number of samples per condition and number of studies have been detailed in Table 1. The inter-study variability parameter represents the amount of variability between the studies considered for meta-analysis. In practice, the observed variability between human data studies is considerable ($\sigma \sim 0.5$) [10]. We chose two different values of $\sigma$ (0.15 and 0.5) to represent small and large amount of inter-study variability respectively. For each setting described in Table 1, 100 independent trials were considered.

| **Setting** | **$\sigma$** | **No. of studies** | **No. of replicates (case, control)** | **AUC (MIN, IN, FIN)** | **Std. dev (MIN, IN, FIN)** |
|---|---|---|---|---|---|
| 1 | 0.15 | 3 | (10, 10) (15, 10) (12, 16) | 0.886, 0.920, 0.920 | 0.005, 0.003, 0.003 |
| 2 | 0.15 | 5 | (10, 10) (15, 10) (12, 16) (14, 12) (20, 20) | 0.953, 0.970, 0.970 | 0.005, < 0.001, 0.001 |
| 3 | 0.5 | 3 | (10, 10) (15, 10) (12, 16) | 0.950, 0.965, 0.966 | 0.004, 0.005, 0.005 |
| 4 | 0.5 | 5 | (10, 10) (15, 10) (12, 16) (14, 12) (20, 20) | 0.957, 0.977, 0.977 | 0.005, 0.005, 0.005 |

**Table 1.** Simulation settings for inter-study variability parameter ($\sigma$), number of studies and number of replicates per study. Area under the ROC curves (AUC) for IN, MIN and FIN methods computed using 100 trials for each simulation setting.

For each simulation setting, individual p-values obtained from differential expression analysis using edgeR were combined using both MIN and IN methods. A gene was considered differentially expressed if the BH adjusted combined p-value (FDR) $< 0.05$. Based on area under the ROC curves (AUC) (Table 1, Figure 1), both meta-analysis methods performed well in terms of detection power in identifying DEGs under all simulation settings. For both low ($\sigma = 0.15$) and high ($\sigma = 0.5$) inter-study variability we observed that the MIN method was more conservative (AUC was smaller than IN) than the IN method. However, as the inter-study variability and the number of studies to be combined increased, both meta-analysis methods were found to have comparable performance (Figure 1c-1d). Although slightly conservative in its performance with respect to the IN approach, MIN method has the advantage of using direction of expression information leading to identification of DEGs among genes with conflicting direction of expression across studies. The conservative behaviour of the MIN method can be attributed to the fact that a two-sided hypothesis testing is performed as compared to a one-sided test on right-hand tail of the distribution in case of IN method. Hence, next we proposed the FIN method as a mixture of IN and MIN methods to circumvent the issue of conservativeness of MIN method.

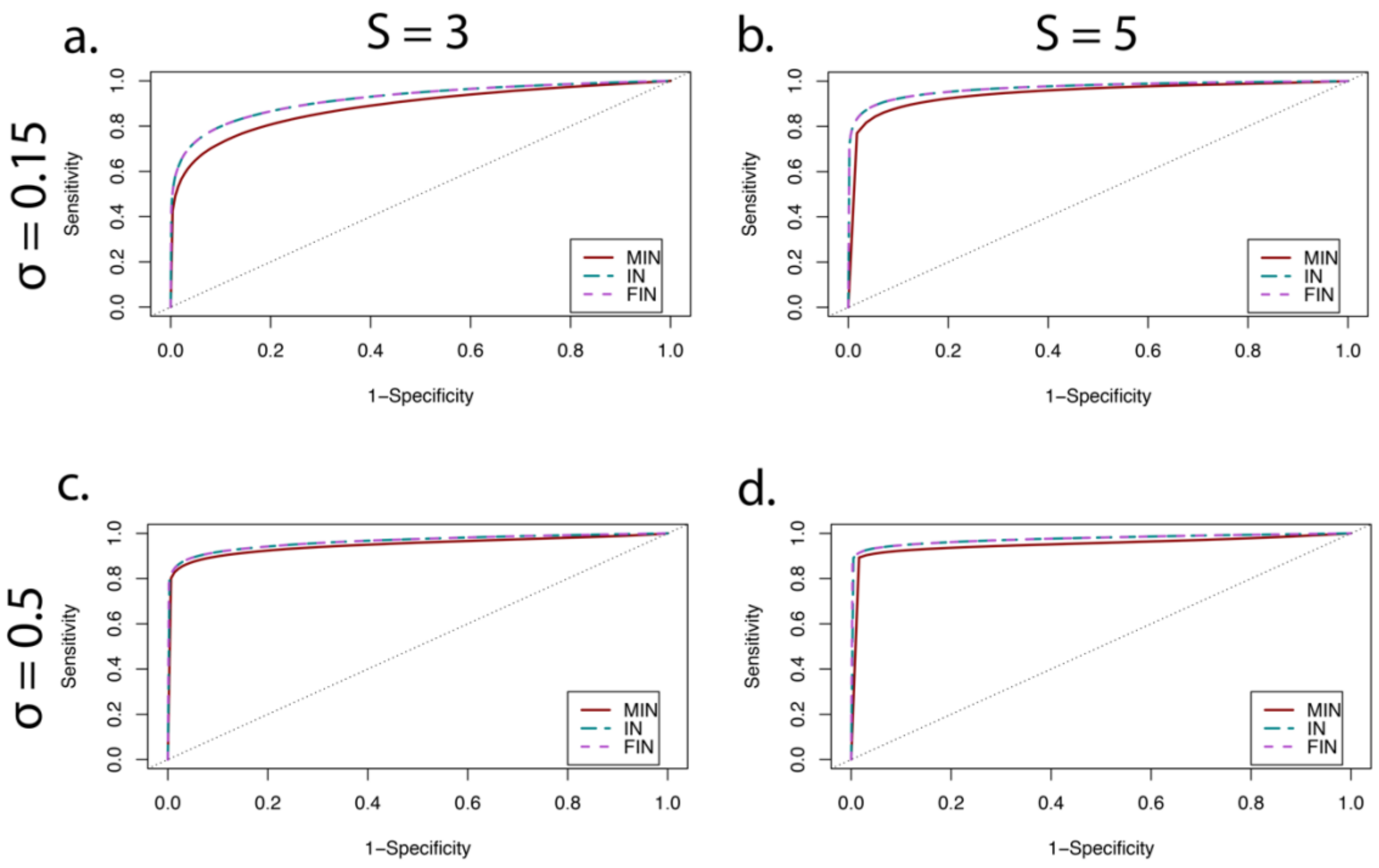

**Figure 1. Performance comparison of modified inverse-normal, inverse-normal and fused inverse-normal methods.** Plots of receiver operating characteristics (ROC) curves averaged over 100 trials for each simulation setting. Simulation settings are represented by rows (from top to bottom): corresponding to low ($\sigma = 0.15$) and high ($\sigma = 0.5$) inter-study variability and columns (from left to right): corresponding to 3 (S=3) and 5 studies (S=5) combined. The red, turquoise and magenta ROC curves represent the modified inverse-normal, inverse-normal and fused inverse-normal methods respectively.

As expected, with increase in inter-study variability and number of studies to be combined, the number of genes with mismatched direction of expression was significantly higher (see Supplementary 1: Table 1). We also note that the FDR for all simulation settings was controlled well below 5% threshold (Figure 2a). In terms of uniquely identified DEGs by the MIN method as compared to IN method, the proportion of true positives (TP) was higher than 80% (Figure 2b) in all simulation settings. A large proportion of TPs among the unique DEGs identified by the MIN method indicates that the MIN approach can lead to DEGs that are biologically relevant to a disease in a real application. Moreover, as the inter-study variability, number of studies or both increased, there was an increase in the number of uniquely identified DEGs by the MIN method and proportion of TPs among them (Figure 2b). More importantly, a high percentage of these unique DEGs (>75% in all settings) were observed to have the true direction of expression (Figure 2c) suggesting that a significantly high percentage of uniquely identified DEGs by the MIN method in real data applications will have true direction of expression as their effective direction of expression. The effective observed direction of expression was determined by the sign of $N_g$ as defined in equation (1).

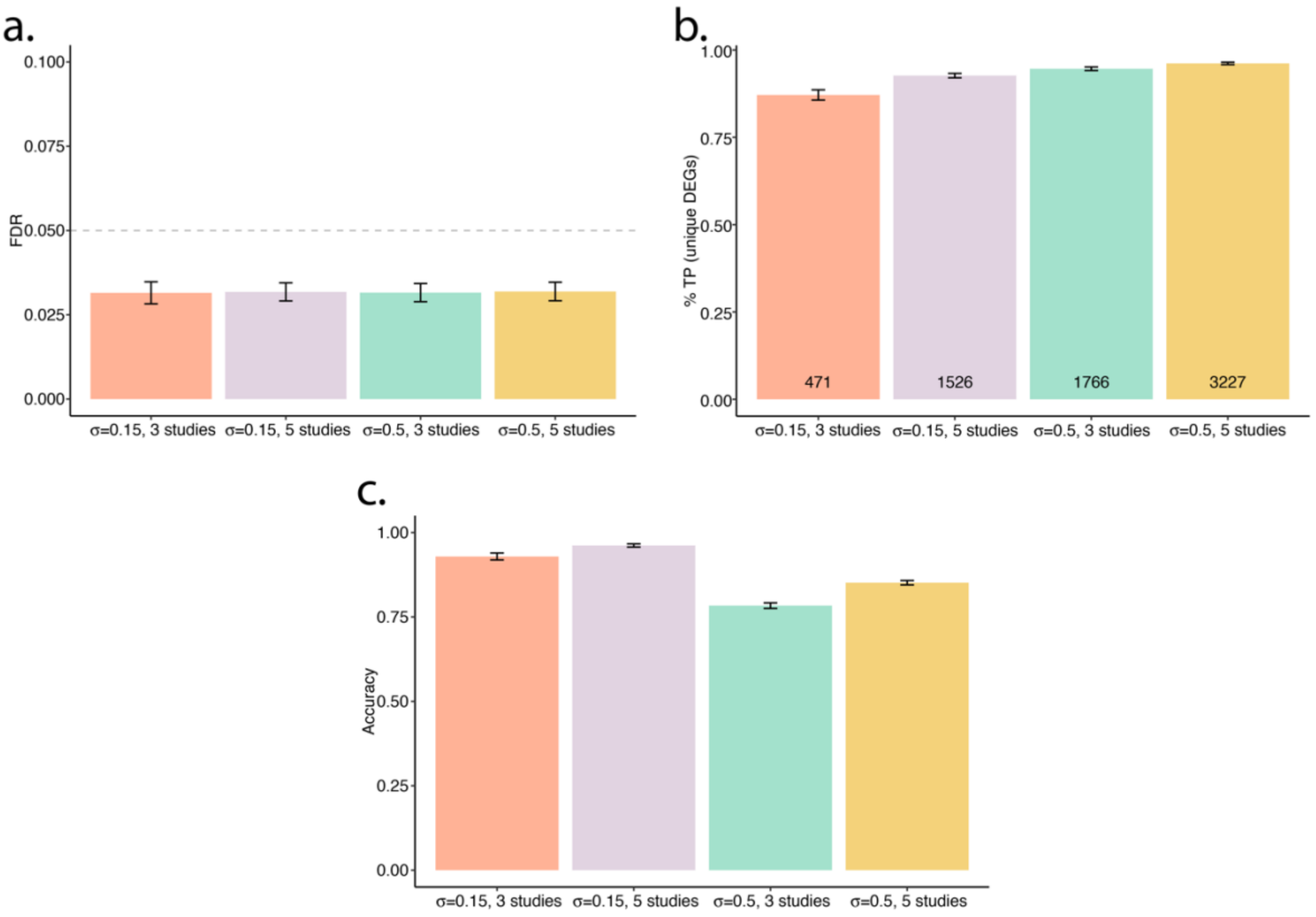

**Figure 2. Characteristics of modified inverse-normal method. a).** False discovery rates for modified inverse-normal method for all simulation settings. **b).** Proportion of true-positives (TPs) among unique differentially expressed genes (DEGs) identified by modified inverse-normal (MIN) method as compared to inverse-normal (IN) method. **c).** Proportion of unique DEGs (MIN) with the observed effective direction of expression as the true direction of expression.

**Fused inverse-normal method**

To address the conservative nature of MIN method, we propose a mixture method which is a mixture of IN and MIN method for integrated differential analysis. In contrast to formula (1) we define $N_g$ as follows:

$$N_g = \begin{cases} \sum_{s=1}^{S} w_s \Phi^{-1}(1 - p_{gs}), & \text{if } g \text{ has same direction of expression across s} \\ \left|\sum_{s=1}^{S} w_s B_{gs} \Phi^{-1}(1 - p_{gs})\right|, & \text{otherwise} \end{cases} \quad (4)$$

Here, $w_s$, $\Phi$ and $B_{gs}$ have their usual meaning as described previously. As $N_g$ follows a standard normal distribution given the assumption that $p_{gs}$ is uniformly distributed under the null hypothesis, a one-sided test on the right-hand tail of the distribution (as proposed in [10]) can be performed for genes with same direction of expression across studies. For the genes with conflicting direction of expression across studies, a two-sided test can be performed. $H_0$ being the same in both the cases. Multiple testing correction to control the overall FDR can then be carried out using the BH method. A detailed interpretation of the FIN model in terms of differential expression of a gene and its direction of expression in individual studies can be found in Supplementary 1.

**Simulation study: FIN, IN and MIN comparison**

In addition to the simulation study for comparing MIN with IN method, we assess and compare the performance of FIN method to that of IN and MIN methods by using the same simulated data and settings described in Table 1. Based on AUC (Table 1, Figure 1), FIN performed similar or better than IN method and had better performance than MIN under all simulation settings. As with MIN, FIN method also has the advantage of using direction of expression information and hence identified DEGs among genes with conflicting direction of expression in contrast with IN method. More importantly, we observed that FIN significantly improved detection power for true DE genes with concordant differential expression patterns across studies as compared to MIN method and does not lead to increased number of false positive detections overall (Figure 3a).

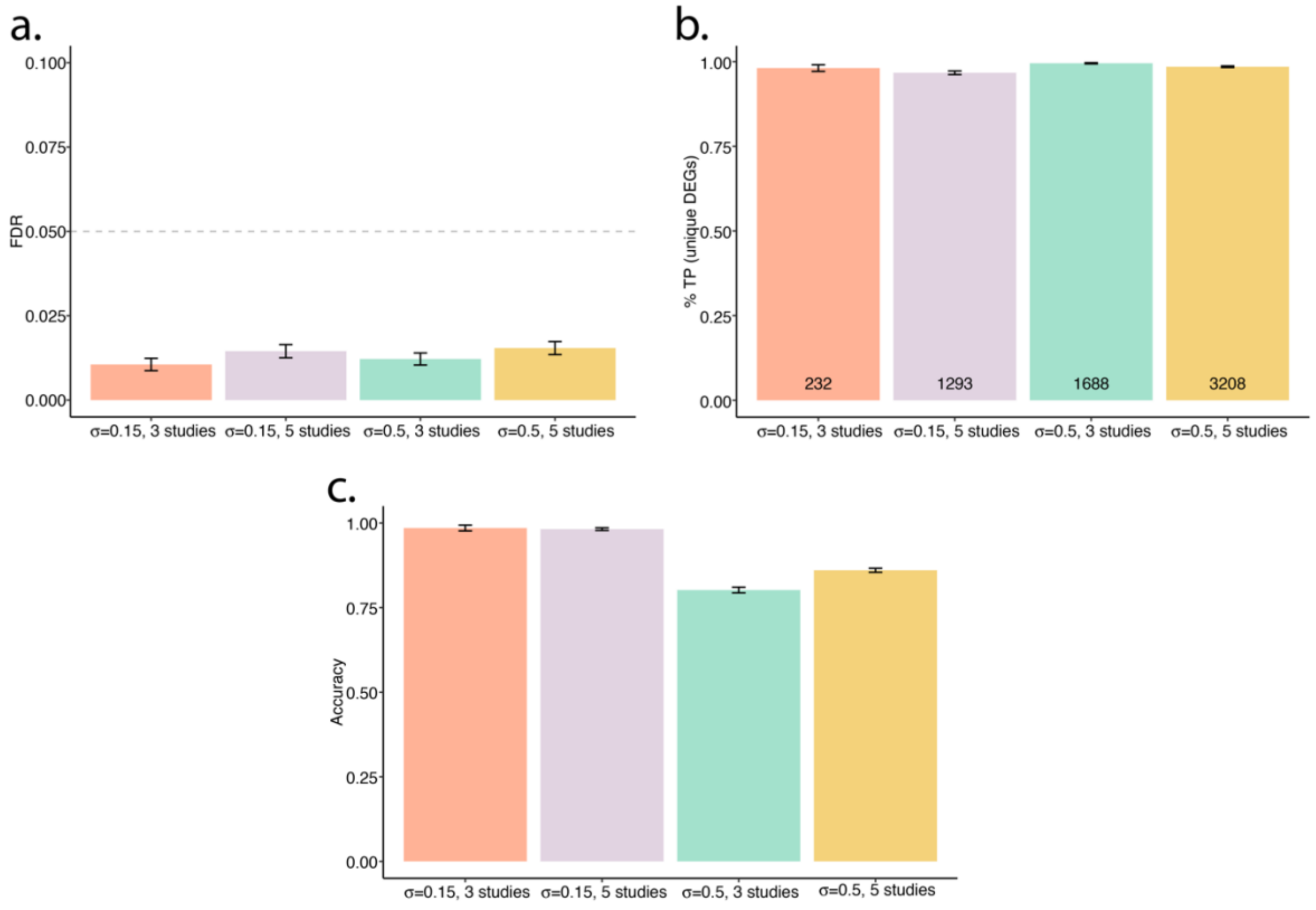

**Figure 3. Characteristics of fused inverse-normal method. a).** False discovery rates for fused inverse-normal method for all simulation settings. **b).** Proportion of true-positives (TPs) among unique differentially expressed genes (DEGs) identified by fused inverse-normal (FIN) method as compared to inverse-normal (IN) method. **c).** Proportion of unique DEGs (FIN) with the observed effective direction of expression as the true direction of expression.

As compared to IN, the proportion of TPs among the uniquely identified DEGs by FIN method was higher than 90% (Figure 3b) indicating that FIN method can lead to DEGs that are biologically relevant to a disease in a real application. Similar to MIN, as the inter-study variability, number of studies or both increased, there was an increase in the number of uniquely identified DEGs by the FIN method as compared to IN method and proportion of TPs among them (Figure 3b). In addition, a high percentage of these unique DEGs (>80% in all settings) were observed to have the true direction of expression (Figure 3c) suggesting that a significantly high percentage of uniquely identified DEGs by the FIN method in real data applications will have true direction of expression as their effective direction of expression. The effective observed direction of expression was determined by the sign of $N_g$ for genes with conflicting direction of expression across studies. In case of same direction of expression of a gene across studies, the consistent direction of expression was kept as the effective direction of expression.

## Application to brain cancer data

To demonstrate how the MIN and FIN method can be adapted in practice for differential meta-analysis of RNA-seq data and compare it with IN method, an application to real GBM studies has been conducted.

### Data description

GBM RNA-seq datasets were searched in GEO (https://www.ncbi.nlm.nih.gov/geo/) and TCGA databases (https://portal.gdc.cancer.gov/). Datasets were selected based on a selection criterion that at least 3 GBM patients and 3 normal tissue samples are available for analysis. Three different GBM RNA-seq datasets, two from GEO (with accession ID: GSE123892 and GSE151352) and one from TCGA (TCGA-GBM) matched our selection criteria and were considered for analysis (for details, see Table 2). Raw gene or transcript counts data (where available) was directly downloaded for TCGA-GBM and GSE123892 datasets. For GSE151352, raw FASTQ files were downloaded and processed using Galaxy web platform via the European UseGalaxy server (https://usegalaxy.eu/) [28] to obtain raw counts. The quality of the raw reads was assessed (using FastQC) and the specified adapter sequence ATCACCGACTGCCCATAGAGAGGCTGAGAC was removed with Cutadapt (version 1.16) [29]. The parameters used for this step were the parameters provided by the submitter of the dataset on GEO. The adapter trimmed reads were aligned to the reference genome (GRCh37.p13) using sequence aligner RNA STAR (Galaxy version 2.7.5b) [30] where other parameters used were default settings. Following alignment, the generated BAM files were processed using the featureCounts tool (Galaxy version 1.6.4+galaxy2) to get raw counts for each RNA-seq data sample. More details of the processing pipeline used for GSE151352 can be found in Supplementary 1: Figure 1.

| Datasets | No. of replicates (Cases/Normal) | No. of genes (after filtering) | Up DEGs | Down DEGs |
|---|---|---|---|---|
| GSE123892 | 4/3 | 15024 | 1914 | 1837 |
| GSE151352 | 12/12 | 12916 | 670 | 1545 |
| TCGA-GBM | 160/5 | 17943 | 3746 | 3183 |

**Table 2.** Information about GBM RNA-seq datasets used for integrated analysis using different p-value combination methods in our study. Up and down DEGs refer to the up and down-regulated DEGs obtained in per-study differential analysis.

### Per-study differential expression analysis

Each of these datasets were processed separately for quality control and differential expression analysis using edgeR package (version 3.26.5) in R. Raw counts data (transcript) were annotated by mapping Ensembl IDs to Entrez Gene IDs and gene symbols (org.Hs.eg.db package, version 3.8.2 in R [31]). Ensembl IDs with no Entrez ID mapping were filtered out. For those with multiple matchings, the one with highest aggregated count was selected. Counts per million (CPM) threshold was carefully selected to reduce the number of low expressed transcripts. We removed a transcript if number of samples equal to or more than the minimum number of samples in a condition had less than 0.85 CPM for that transcript. Although subjective, this choice of threshold seems to work well for the uniform distribution assumption for the p-values under $H_0$. Only genes left after low expression filtering were considered for individual differential expression analysis in order to satisfy the uniformity assumption on p-values under the null hypothesis (Figure 4a).

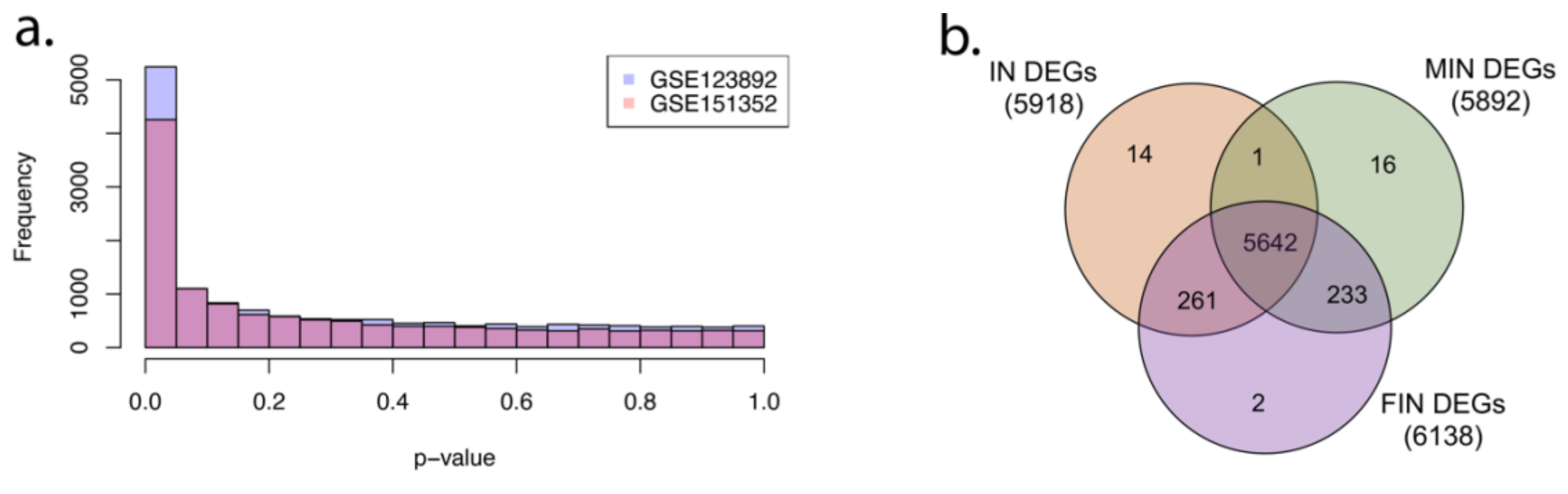

**Figure 4. Comparison of results from meta-analysis methods. a).** Histograms of raw p-values obtained from per-study differential analysis of GSE123892 and GSE151352 GBM datasets from gene expression omnibus used in real data application. **b).** Venn diagram of the differentially expressed genes identified using inverse-normal (IN DEGs), modified inverse-normal (MIN DEGs) and fused inverse-normal (FIN DEGs) methods.

The remaining transcripts were then normalized using the TMM method [32]. Common and tag-wise dispersion were estimated and a negative-binomial generalized log-linear model was fitted to the read counts using the glmFit function under the edgeR package. Raw p-values were then obtained from the differential analysis for case/control conditions. More information about the DEGs identified in per-study differential analysis based on the criteria $|\log_2 FC| > 1$ and FDR p-value $< 0.05$ can be found in Table 2. We note that TCGA-GBM dataset has a much larger library size (~ 47 million reads, Illumina HiSeq 2000 v2 sequencer) as compared to GSE151352 (~ 4 million reads, Ion Torrent S5 sequencer) and GSE123892 (~ 35 million reads, Illumina HiSeq 2500 sequencer). Hence, we observed a differing number of genes left after filtering and consequently a much larger number of DEGs being observed for TCGA-GBM dataset as compared to the other two (Table 2) [33] in per-study differential expression analysis. As the sequencing output gets larger, the smaller count differences between samples are declared significant by models for differential expression in edgeR. A more detailed treatment of differential expression in RNA-seq data and how it is affected by sequencing depth and other factors can be found in Tarazona et al. (2011) [33].

Moreover, we also considered individual differential analysis for TCGA-GBM RNA-seq data by randomly selecting 20 cases together with available 5 normal samples in order to make all three datasets (GSE123892, GSE151352 and TCGA-GBM) comparable in terms of number of replicates for the meta-analysis (see Supplementary 1: Table 2). Hence, we considered two different meta-analysis scenarios.

**Meta-analysis**

Once the raw p-values were obtained from the individual differential expression analysis for each dataset, IN, MIN and FIN methods were then applied for p-value combination. Since the TCGA-GBM dataset (160 GBM vs 5 normal samples) is much larger in terms of number of samples as compared to GSE123892 (4 GBM vs 3 normal samples) and GSE151352 (12 GBM vs 12 normal samples), we considered two different combination scenarios. First, all TCGA-GBM samples were used for individual analysis to obtain the raw p-values. Second, 20 cases and 5 normal samples randomly selected from TCGA-GBM dataset were considered for individual analysis to get raw p-values and then considered for meta-analysis with the other two datasets (GSE123892 and GSE151352). 10 different random selections were made and individual differential expression analysis were conducted respectively. Second scenario ensured that the datasets included in meta-analysis had comparable sample sizes.

In scenario one, a total of 18325 unique gene pool was considered for meta-analysis which was the combination of genes identified in each RNA-seq data analysis after quality control and filtering (Table 2). 13056 out of 18325 genes (~71%) were found to have the same direction of expression across the studies in which they were present whereas 5259 (~29%) of genes had conflicting or mismatched direction of expression. The direction of expression for a gene in an individual study was determined based on the sign of $\log_2 FC$ obtained for that gene in per-study differential analysis. Hence, only 13056 genes were effectively considered for IN method as compared to 18325 genes for MIN and FIN methods for identifying DEGs because of post-hoc removal of DEGs with conflicting direction of expression in the IN method.

For each of the combination methods, we assessed the number of DEGs based on average absolute log fold-change $\sum_{i=1}^{n}|\log_2 FC_i|/n > 1$ and FDR p-value $< 0.05$ criteria. Here, $n$ denotes the number of datasets in which a particular gene was present. In case a gene was absent in a dataset, the weights in the combination methods were estimated only using the number of replicates in datasets in which the gene was present. The three p-value combination methods were then compared based on number of DEGs identified and unique DEGs identified by each method. A total of 5918, 5892 and 6138 DEGs were identified by the IN, MIN and FIN methods respectively. Of the DEGs detected by all these meta-analysis methods, more than 90% of them were in common (Figure 4b) with FIN method having a higher detection power than the other two methods. Moreover, MIN and FIN method identified a total of 233 and 235 DEGs with mismatched direction of expression across studies by incorporating the direction of expression information. More importantly, in the subset of DEGs which were present in all three datasets, 5.26% of DEGs had conflicting direction of expression across studies. Although, small in proportion, this would be of importance in case a gene of interest for the disease being studied has conflicting direction of expression across different studies. Particularly when more datasets are included in meta-analysis, the number of genes considered in IN approach can be massively reduced.

Given that the FIN method has the highest power of DEG detection, we further explore the DEGs obtained using this meta-analysis procedure. Top 10 up and down-regulated DEGs identified by FIN method are presented in Table 3. For full list of DEGs identified by different meta-analysis methods, see Supplementary 2, 3 and 4. In terms of effective direction of expression of DEGs, 2914 DEGs with same direction of expression across studies and 180 DEGs with mismatched direction of expression were up-regulated. Similarly, 2989 (same direction) and 55 (mismatched direction) DEGs were down-regulated. Results for scenario 2 for random selection have been detailed in Supplementary 1: Table 3, 4 and 5.

| **DEGs (Up)** | $N_g$ | **Mean \|logFC\|** | **Effect** | **BH p-value** | **DEGs (Down)** | $N_g$ | **Mean \|logFC\|** | **Effect** | **BH p-value** |
|---|---|---|---|---|---|---|---|---|---|
| *EIF4EBP1* | 10.45 | 3.33 | +++ | $< 1.62 \times 10^{-15}$ | *SMAD12* | 11.19 | 4.32 | --- | $< 1.62 \times 10^{-15}$ |
| *WEE1* | 10.39 | 4.04 | +++ | $< 1.62 \times 10^{-15}$ | *RASGRF2* | 11.10 | 4.19 | --- | $< 1.62 \times 10^{-15}$ |
| *VIM* | 10.39 | 3.68 | +++ | $< 1.62 \times 10^{-15}$ | *DNAJC6* | 11.07 | 3.71 | --- | $< 1.62 \times 10^{-15}$ |
| *NUSAP1* | 10.29 | 4.67 | +++ | $< 1.62 \times 10^{-15}$ | *SERPINI1* | 10.99 | 4.79 | --- | $< 1.62 \times 10^{-15}$ |
| *HJURP* | 10.24 | 5.79 | +++ | $< 1.62 \times 10^{-15}$ | *ATP1B1* | 10.98 | 3.35 | --- | $< 1.62 \times 10^{-15}$ |
| *KIF4A* | 10.15 | 4.48 | +++ | $< 1.62 \times 10^{-15}$ | *ATP8A1* | 10.91 | 3.95 | --- | $< 1.62 \times 10^{-15}$ |
| *KIF20A* | 10.12 | 5.80 | +++ | $< 1.62 \times 10^{-15}$ | *JAKMIP3* | 10.91 | 4.40 | --- | $< 1.62 \times 10^{-15}$ |
| *AURKB* | 10.09 | 5.48 | +++ | $< 1.62 \times 10^{-15}$ | *MFSD6* | 10.90 | 2.83 | --- | $< 1.62 \times 10^{-15}$ |
| *UBE2C* | 10.07 | 5.95 | +++ | $< 1.62 \times 10^{-15}$ | *DCTN1-AS1* | 10.88 | 5.33 | --- | $< 1.62 \times 10^{-15}$ |
| *CCNB2* | 10.04 | 4.63 | +++ | $< 1.62 \times 10^{-15}$ | *PRKACB* | 10.85 | 2.35 | --- | $< 1.62 \times 10^{-15}$ |

**Table 3.** Top 10 up- and down-regulated DEGs identified by FIN method. The DEGs have been sorted based on the value of the statistic $N_g$ and the mean of absolute value of the $\log_2 FC$ have been reported. Effect signifies the direction of expression of DEGs in the per-study differential analysis.

**Pathway analysis and biological significance**

DEGs obtained by the FIN method were further explored to assess their biological relevance to GBM. QIAGEN's Ingenuity Pathway Analysis (IPA) (www.qiagen.com/ingenuity) tool was used to identify biological pathways in which DEGs were enriched and upstream regulator analysis (URA) identified upstream regulators for GBM. We performed pathway analysis and URA separately for DEGs that were up-regulated and down-regulated and were present in all three datasets. We also note that not all identified DEGs by the meta-analysis methods are present in all three studies considered. Number of DEGs present in one, two or all three datasets have been detailed in Table 4.

| Method | Expression direction | Present in one study | Present in two studies | Present in three studies | Total DEGs |
|---|---|---|---|---|---|
| IN | Same | 1368 | 1085 | 3465 | 5918 |
| | Mismatched | 0 | 0 | 0 | |
| MIN | Same | 1182 | 1035 | 3442 | 5892 |
| | Mismatched | 0 | 52 | 181 | |
| FIN | Same | 1359 | 1083 | 3461 | 6138 |
| | Mismatched | 0 | 53 | 182 | |

**Table 4.** Number of DEGs found in one, two or all three datasets. Same and mismatched represents if the direction of expression of a DEG was consistent across a study or not respectively.

Of 1798 up-regulated DEGs, all of them mapped in the IPA database and 128 canonical pathways were identified based on BH adjusted p-value (< 0.01). These include Heptatic Fibrosis Signaling Pathway (adj. Pval. $= 3.98 \times 10^{-13}$, ratio = 0.205), Kinetochore Metaphase Signaling Pathway (adj. Pval. $= 7.94 \times 10^{-15}$, ratio = 0.376), Cell Cycle Control of Chromosomal Replication (adj. Pval. $= 5.89 \times 10^{-09}$, ratio = 0.393), Role of BRCA1 in DNA Damage Response (adj. Pval. $= 1.32 \times 10^{-08}$, ratio = 0.325) and IL-8 Signaling (adj. Pval. $= 4.37 \times 10^{-08}$, ratio = 0.215) as some of the top dysregulated pathways. More importantly, major aberrant pathways shown to be involved in GBM pathogenesis [34][35] were also identified and include Glioblastoma Multiforme Signaling (adj. Pval. $= 2.95 \times 10^{-06}$, ratio = 0.206), Glioma Signaling (adj. Pval. $= 3.63 \times 10^{-05}$, ratio = 0.205), p53 Signaling (adj. Pval. $= 5.25 \times 10^{-05}$, ratio = 0.224), Glioma Invasiveness Signaling (adj. Pval. $= 0.0008$, ratio = 0.219), PI3K/AKT Signaling (adj. Pval. $= 0.005$, ratio = 0.146) and mTOR Signaling (adj. Pval. $= 0.007$, ratio = 0.138).

Similarly, all 1845 down-regulated DEGs mapped to the IPA database and 98 canonical pathways were identified as significant (BH adjusted p-value < 0.01). Synaptogenesis Signaling Pathway (adj. Pval. $= 3.16 \times 10^{-25}$, ratio = 0.288), Endocannabinoid Neuronal Synapse Pathway (adj. Pval. $= 1.00 \times 10^{-16}$, ratio = 0.359), Opioid Signaling Pathway (adj. Pval. $= 1.00 \times 10^{-13}$, ratio = 0.247), GNRH Signaling (adj. Pval. $= 6.31 \times 10^{-11}$, ratio = 0.260), Calcium Signaling (adj. Pval. $= 6.31 \times 10^{-11}$, ratio = 0.243), G Beta Gamma

Signaling (adj. Pval. = $9.33 \times 10^{-10}$, ratio = 0.287) and Dopamine-DARPP32 Feedback in cAMP Signaling (adj. Pval. = $1.55 \times 10^{-09}$, ratio = 0.252) were identified as some of the top dysregulated pathways. The top 10 pathways identified by the up-regulated and down-regulated DEGs separately are illustrated in Figure 5. For complete list of identified pathways for up- and down-regulated DEGs in our study, see Supplementary 1: Table 6.

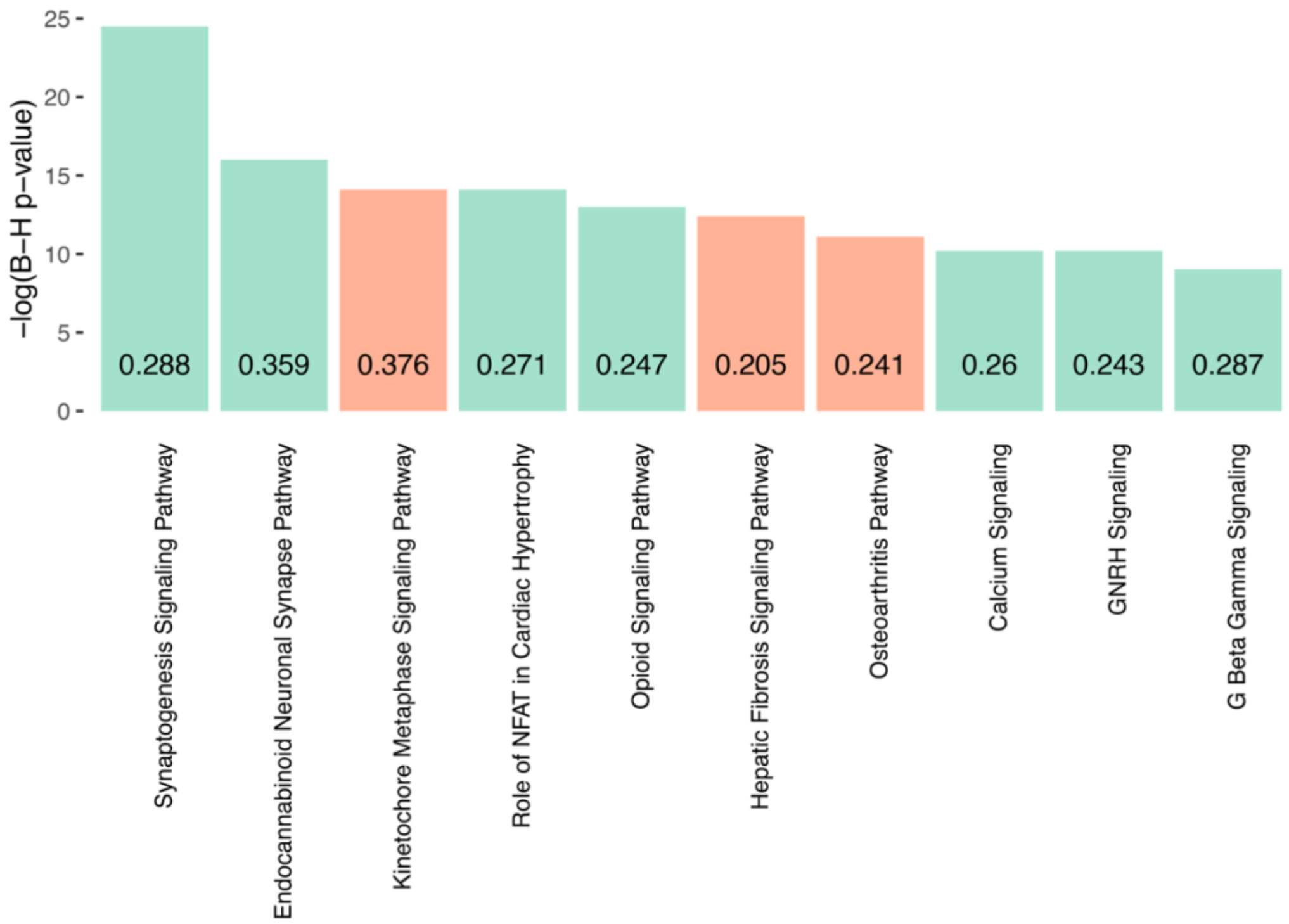

**Figure 5. Significant pathways identified by IPA.** The top ten significant pathways based on BH p-value among the canonical pathways identified by IPA for the up-regulated DEGs (orange bar) and down-regulated DEGs (green bar). The numbers on the bar plot show the ratio between the numbers of DEGs enriched and total number of genes in each of these pathways.

In addition, the URA tool in IPA identified potential upstream regulators (transcription factors, genes or other small molecules) that has been experimentally observed to affect gene expression. It identifies these regulators by analysing linkage to DEGs through coordinated expression [36]. Among the up-regulated DEGs, *TGFB1* and *TP53*, which are also DEGs and important in GBM pathogenesis [37][38] are predicted to be the top two upstream regulators. 293 up-regulated DEGs were identified as potential upstream regulators of gene upregulation out of a total of 2215 (BH corrected p-value $< 0.01$, see Supplementary 1: Table 7a) predicted upstream regulators. Out of 2215 predicted, 764 of these significant upstream regulators were activated and 112 were also observed as DEGs in our analysis.

On the contrary, for the down-regulated DEGs, IPA identified 32 potential upstream regulators (BH corrected p-value $< 0.01$, see Supplementary 1: Table 7b) with *TCF7L2* and *MAPT* as the top two. 14 of the 32 upstream regulated were predicted to be inhibited and two among the inhibited are DEGs. *TCF7L2* is a diabetes risk-associated gene which plays a key role in the Wnt-signaling pathway and is shown to be frequently mutated in colorectal cancer [39] and promote cell proliferation [40]. However, exploration of its role in GBM pathogenesis warrant further studies. Interestingly, *MAPT* is also a DEG observed in our analysis and is one of the two hallmarks of AD [41]. Gargini et al. (2019) [42] observed a strong correlation of Tau/MAPT expression and indicators of survival in glioma patients. Moreover, it has been found to be epigenetically controlled by balance between *IDH1/2* wild-type and mutation in human gliomas [43]. Thus, providing further evidence and reaffirming the involvement of *MAPT* in central nervous system disorders.

Of the DEGs with conflicting direction of expression across studies, 182 out of 235 DEGs are present in all three datasets. Among them *CMTM6, RAD51*, *NOS1AP*, *MSANTD1*, *PGM2, PSD3*, *GPR82*, *SPTBN4*, *TSPAN6* and *ARHGEF28* were identified as top 10 DEGs based on the absolute value of $N_g$ (see Table 5). Interestingly, *RAD51* and *ARHGEF28* have previously been identified as a tumour suppressor and an oncogene respectively [44]. More importantly, *RAD51* was found to be effectively overexpressed in GBM in our study and have recently been shown as a target for inhibition to enhance radiosensitivity of GBM cells during treatment [45][46]. On the other hand, *ARHGEF28* was found to be effectively down-regulated in our study. It is an intracellular kinase that functions either as a Rho guanine exchange factor or a scaffolding protein to initiate FAK activation and cell contractibility [47]. Furthermore, the RhoA-FAK pathway has been shown to be involved in colon cancer cell proliferation and migration [48]. *ARHGEF28* mRNA levels have also been found to be elevated in late-stage ovarian cancer and associated with decreased progression free and overall survival [49]. However, its role in GBM growth and progression is yet to be elucidated and requires exploration in future studies.

| DEGs | $N_g$ | Mean \|logFC\| | Effect | BH p-value |
|---|---|---|---|---|
| *CMTM6* | 7.58 | 1.30 | +-+ | $3.93 \times 10^{-13}$ |
| *RAD51* | 7.58 | 2.82 | +-+ | $4.03 \times 10^{-13}$ |
| *NOS1AP* | -7.53 | 1.37 | -+- | $5.73 \times 10^{-13}$ |
| *MSANTD1* | -7.53 | 1.31 | -+- | $5.92 \times 10^{-13}$ |
| *PGM2* | 7.52 | 1.31 | +-+ | $6.35 \times 10^{-13}$ |
| *PSD3* | -7.47 | 1.67 | -+- | $8.63 \times 10^{-13}$ |
| *GPR82* | 7.43 | 4.35 | +-+ | $1.24 \times 10^{-12}$ |
| *SPTBN4* | -7.31 | 1.91 | -+- | $2.78 \times 10^{-12}$ |
| *TSPAN6* | 7.18 | 2.08 | +-+ | $6.86 \times 10^{-12}$ |
| *ARHGEF28* | -7.06 | 1.23 | +-- | $1.63 \times 10^{-11}$ |

**Table 5.** Top 10 DEGs with mismatched direction of expression across datasets identified by FIN method. The DEGs have been sorted based on the absolute value of the statistic $N_g$ and the mean of absolute value of the $\log_2 FC$ have been reported. Effect signifies the direction of expression of DEGs in the per-study differential analysis for GSE123892, GSE151352 and TCGA-GBM respectively.

## Discussion

Although the implementation of MIN and FIN p-value combination methods are straight forward, they require some additional considerations. First, the used weighting criteria leads to a larger weight being given to a study with larger sample sizes. Intuitively, this is expected as a study with a larger sample size might be more robust than studies with lower sample sizes. However, importance must also be given to the quality of the RNA-seq data in each study. It must be assessed in case this information is available and other weights more appropriate as per the quality of the data may be specified. In our study, since TCGA-GBM has a much larger sample size as compared to the other two datasets, we compared the number of DEGs obtained by FIN method by considering all 165 samples and 10 random selections of 20 cases and 5 normal samples. On average, about 94% of the DEGs obtained when randomly selected subset was considered were also found in DEGs identified using the full TCGA-GBM dataset (see

Supplementary 1: Table 8). Hence, suggesting that the identification of DEGs was stable across these two settings.

Next, the MIN and FIN are adaptive in a sense that they allow for consideration of genes that may not be present in all studies that are considered for integrated differential expression analysis. In case a gene is not present in some of the studies, the weights ($w_s$) in the combination method can only be estimated using the number of replicates in the datasets in which the gene is present. However, for genes that are just present in one study, it would mean that the results from the meta-analysis for these genes would be the same as the per-study differential analysis. Hence, a careful consideration about the quality of the RNA-seq data and library size is required in case only the genes that are common among studies are considered. For datasets of similar quality and library size, a large proportion of genes would not be excluded from meta-analysis if only common genes are used. However, a large number of genes might be excluded from meta-analysis in case of dissimilar library sizes and quality which could lead to potentially missing out on important genes for the disease. For instance, only 12345 out of 18315 unique genes are present in all 3 studies in our application where the library sizes are not similar. Thus, a balanced approach is suggested.

Finally, we used edgeR for per-study differential analysis in our study but other popular packages such as DESeq2 [50] and NOIseq [33] can be applied. Moreover, the FIN model can be extended to multi-group comparisons apart from a two-group comparison discussed in this study. The proposed meta-analysis method relies on the fact that the same test statistics are used for per-study differential expression analysis to obtain individual p-values and all studies under consideration have the same experimental considerations. For instance, in case DESeq2 is used for multi-group differential expression analysis in each study, a likelihood ratio test is used rather than Wald statistics being used for two group differential expression analysis.

## Conclusions

In this study, we proposed MIN and consequently FIN method for meta-analysis of RNA-seq data. The developed methods account for both the sample size of study and direction of expression of a gene in each study allowing for detection of potentially robust biologically significant DEGs even when they have conflicting direction of expression across studies. In contrast with the existing IN method, the proposed methods have the advantage of identifying DEGs among genes with conflicting direction of expression across studies. For the genes with concordant differential expression patterns across studies the MIN method exhibited a similar DEG detection power and performance as compared to IN method particularly when there was high inter-study variability and increased number of studies were considered. FIN method exhibited a similar or improved DEG detection power as compared to IN method and was significantly better in performance as compared to MIN method. More importantly, in a real data application, we demonstrated the use of FIN method in detection of biologically relevant DEGs to GBM. Hence, this meta-analysis method provides a way to establish differential expression status for genes with conflicting direction of expression in individual RNA-seq studies and further exploration of them as potential biomarkers for the disease. With lowering costs and increase in the number of RNA-seq studies being archived on public databases, this method might provide a way to integrate a greater number of studies without losing much prior information and consequently considering all the genes in the analysis irrespective of their direction of expression.

## Declarations

### Ethics approval and consent to participate

Not applicable

### Consent for publication

Not applicable

### Availability of data and materials

All datasets used in study are publicly accessible and appropriate access links have been provided in the text of the manuscript when they have been first mentioned. An implementation of the proposed method can be found here:
https://github.com/nash5202/FIN_meta_analysis/

### Competing interests

The authors declare that they have no competing interests.

### Funding

This research was supported by AiPBAND (www.aipband-itn.eu), European Union's Horizon 2020 research and innovation programme under the Marie Skłodowska-Curie grant agreement 764281. BP is a Marie-Curie early-stage research fellow of AiPBAND.

### Authors' contributions

BP and XL conceived and designed the study. BP performed simulations, data analysis and interpretation and drafted the manuscript. XL helped interpret the results, reviewed and edited the manuscript. All authors read and approved the final manuscript.

### Authors' information

[1]National Horizons Centre, School of Health and Life Sciences, Teesside University, Darlington, DL1 1HG, UK.
*Correspondence: Email: X.Li@tees.ac.uk, Phone: +44-01642738451

## Figures, tables and additional files

### Figures

**Figure 3. Performance comparison of modified inverse-normal, inverse-normal and fused inverse-normal methods.** Plots of receiver operating characteristics (ROC) curves averaged over 100 trials for each simulation setting. Simulation settings are represented by rows (from top to bottom): corresponding to low ($\sigma = 0.15$) and high ($\sigma = 0.5$) inter-study variability and columns (from left to right): corresponding to 3 (S=3) and 5 studies (S=5) combined. The red, turquoise and magenta ROC curves represent the modified inverse-normal, inverse-normal and fused inverse-normal methods respectively.

**Figure 4. Characteristics of modified inverse-normal method. a).** False discovery rates for modified inverse-normal method for all simulation settings. **b).** Proportion of true-positives (TPs) among unique differentially expressed genes (DEGs) identified by modified inverse-normal (MIN) method as compared to inverse-normal (IN) method. **c).** Proportion of unique DEGs (MIN) with the observed effective direction of expression as the true direction of expression.

**Figure 3. Characteristics of fused inverse-normal method. a).** False discovery rates for fused inverse-normal method for all simulation settings. **b).** Proportion of true-positives (TPs) among unique differentially expressed genes (DEGs) identified by fused inverse-normal (FIN) method as compared to inverse-normal (IN) method. **c).** Proportion of unique DEGs (FIN) with the observed effective direction of expression as the true direction of expression.

**Figure 4. Comparison of results from meta-analysis methods. a).** Histograms of raw p-values obtained from per-study differential analysis of GSE123892 and GSE151352 GBM

datasets from gene expression omnibus used in real data application. **b).** Venn diagram of the differentially expressed genes identified using inverse-normal (IN DEGs), modified inverse-normal (MIN DEGs) and fused inverse-normal (FIN DEGs) methods.

**Figure 5. Significant pathways identified by IPA.** The top ten significant pathways based on BH p-value among the canonical pathways identified by IPA for the up-regulated DEGs (orange bar) and down-regulated DEGs (green bar). The numbers on the bar plot show the ratio between the numbers of DEGs enriched and total number of genes in each of these pathways.

**Tables**

**Table 1.** Simulation settings for inter-study variability parameter ($\sigma$), number of studies and number of replicates per study. Area under the ROC curves (AUC) for IN, MIN and FIN methods computed using 100 trials for each simulation setting.

**Table 2.** Information about GBM RNA-seq datasets used for integrated analysis using different p-value combination methods in our study. Up and down DEGs refer to the up and down-regulated DEGs obtained in per-study differential analysis.

**Table 3.** Top 10 up- and down-regulated DEGs identified by FIN method. The DEGs have been sorted based on the value of the statistic $N_g$ and the mean of absolute value of the $\log_2 FC$ have been reported. Effect signifies the direction of expression of DEGs in the per-study differential analysis.

**Table 4.** Number of DEGs found in one, two or all three datasets. Same and mismatched represents if the direction of expression of a DEG was consistent across a study or not respectively.

**Table 5.** Top 10 DEGs with mismatched direction of expression across datasets identified by FIN method. The DEGs have been sorted based on the absolute value of the statistic $N_g$ and the mean of absolute value of the $\log_2 FC$ have been reported. Effect signifies the direction of expression of DEGs in the per-study differential analysis for GSE123892, GSE151352 and TCGA-GBM respectively.

**Supplementary Information**

**Supplementary 1.** Document contains interpretation of the FIN model, details of RNA-seq raw data processing using GALAXY, brief description of the simulation method, results of scenario 2 considered for meta-analysis and Ingenuity pathway analysis results.

**Supplementary 2.** Full list of DEGs identified in scenario 1 by the inverse-normal method in GBM data application.

**Supplementary 3.** Full list of DEGs identified in scenario 1 by the modified inverse-normal method in GBM data application.

**Supplementary 4.** Full list of DEGs identified in scenario 1 by the fused inverse-normal method in GBM data application.

# Fused inverse-normal method for integrated differential expression analysis of RNA-seq data

Birbal Prasad[1] and Xinzhong Li[1*]

[1]National Horizons Centre, School of Health and Life Sciences, Teesside University, Darlington, DL1 1HG, UK. Email: B.Prasad@tees.ac.uk; X.Li@tees.ac.uk

*Correspondence: Email: X.Li@tees.ac.uk, Phone: +44-01642738451

## Supplementary information contents lists:

1. **Interpretations of the FIN model**
2. **Simulation study model**
3. **Processing of raw RNA-seq dataset GSE151352 using GALAXY**
4. **Supplementary Figure 1.** Flow of different steps used in processing of the raw RNA-seq fastq files for GSE151352 using Galaxy to get raw counts.
5. **Supplementary Table 1.** Total number of genes (# genes), number of genes with conflicting direction of expression (# conf. genes), number of differentially expressed genes (# DEGs), number of genes with conflicting direction of expression (# conf. DEGs) and number of true DEGs (# true DEGs) for all simulation settings averaged over 100 trials.
6. **Supplementary Table 2.** Total number of up and down-regulated DEGs identified in per-study differential analysis in 10 random selections of 20 GBM cases and 5 controls from TCGA-GBM dataset.
7. **Supplementary Table 3.** Number of DEGs identified by the inverse-normal (IN) method in the GBM data application for both scenarios and number of these DEGs present in one, two or all three datasets considered.
8. **Supplementary Table 4.** Number of DEGs identified by the modified inverse-normal (MIN) method in the GBM data application for both scenarios and number of these DEGs present in one, two or all three datasets considered.
9. **Supplementary Table 5.** Number of DEGs identified by the fused inverse-normal (FIN) method in the GBM data application for both scenarios and number of these DEGs present in one, two or all three datasets considered.

10. **Supplementary Table 6.** IPA canonical pathways for up and down-regulated DEGs present in all three GBM datasets and identified by the FIN method.

11. **Supplementary Table 7.** IPA upstream regulator analysis results for up and down-regulated DEGs present in all three GBM datasets and identified by the FIN method.

12. **Supplementary Table 8.** Number of common DEGs identified by the FIN method in both scenarios considered for TCGA-GBM data. On average about 94% of the DEGs obtained when randomly selected subset was considered were also found in DEGs identified using the full TCGA-GBM dataset.

13. **Supplementary 2.** Full list of DEGs identified in scenario 1 by the inverse-normal method in GBM data application (in an Excel file).

14. **Supplementary 3.** Full list of DEGs identified in scenario 1 by the modified inverse-normal method in GBM data application (in an Excel file).

15. **Supplementary 4.** Full list of DEGs identified in scenario 1 by the fused inverse-normal method in GBM data application (in an Excel file).

**Interpretations of the FIN model**

In order to understand the practical behaviour and interpretations of the FIN model in different scenarios, we consider the following cases in terms of differential expression of a gene and its direction of expression (up: +, down: -) in individual studies:

i. A gene is non differentially expressed (NDE) in one study and differentially expressed (DE) in another: If it is NDE, $\Phi^{-1}(1-p_{gs})$ is standard normal distributed. Hence, it takes values around 0 with high probability. On the contrary, if it is DE, $\Phi^{-1}(1-p_{gs})$ will take extreme values. Now, the value of $N_g$ will then depend on which study has the larger weight ($w_s$).

ii. A gene is strongly +DE in one study and strongly -DE in another: In this case, $\Phi^{-1}(1-p_{gs})$ will be extreme for both studies. Then, if the weight $w_s$ is similar for both studies, the contribution of the study specific term to $N_g$ will cancel each other out because of $B_{gs}$. Hence, $N_g$ which will be close to 0 will give us non-significant p-value in hypothesis testing for that gene. This makes sense as we will have comparable evidence of conflicting direction of expression for the gene. In case, the weight of one study is comparatively larger than the other study, we will get that the gene is DE with the effective sign of regulation of that of the bigger study.

iii. A gene is strongly +DE in most studies and weakly -DE in a few: Similar to case ii, $\Phi^{-1}(1-p_{gs})$ will be extreme for all the studies. For comparable weights of studies, we would get a large positive value for $N_g$ as we have many more studies where the gene is +DE as compared to where the gene is -DE. Hence, resulting in the gene to be +DE after hypothesis testing. Only in case when the study where the gene is -DE has an extremely larger weight $w_s$ as compared to all other four studies combined can that gene be -DE as a result of the hypothesis testing.

## Simulation study model

Here, we briefly describe the theoretical framework of the simulation study method adapted from Rau et al. (2014) [reference 10 in the manuscript]. For detailed procedure for estimation of parameter values based on real RNA-seq datasets (GSE125583 in this study), see Rau et al. (2014).

Let $Y_{gcrs}$ be random variable with $y_{gcrs}$ as its realisations which are observed count for a gene $g$, condition $c$, biological replicate $r$ and study $s$. RNA-seq data was generated as per the negative binomial distribution,

$$Y_{gcrs} \sim NB(\mu_{gcs}, \phi_{gs})$$

where parameters $\mu_{gcs}$ and $\phi_{gs}$ represent the mean and dispersion, respectively, for a gene $g$, condition $c$ and study $s$. To account for variability between the individual studies (inter-study variability) considered for meta-analysis, we consider the following situation:

$$\mu_{gcs} = \theta_{gc} \times e^{\epsilon_{gcs}}$$

where $\epsilon_{gcs} \sim N(0, \sigma^2)$. $\theta_{gc}$ is the mean for a gene $g$ in condition $c$. $\epsilon_{gcs}$ represents the variability around $\theta_{gc}$ due to a study and condition specific random effect. $\sigma^2$ represents the size of the inter-study variability which affects $\mu_{gcs}$ through $\epsilon_{gcs}$ with $\epsilon_{gcs}$ having a multiplicative effect on $\mu_{gcs}$.

## Processing of raw RNA-seq dataset GSE151352 using GALAXY

Raw fastq files generated by GPL23934 (Ion Torrent S5 (Homo sapiens)) platform were retrieved for all 24 samples (12 tumour and 12 healthy) from Sequence Read Archive (https://trace.ncbi.nlm.nih.gov/Traces/sra/sra.cgi?study=SRP265074) using Get Data tool in Galaxy. Next, we assessed the quality of the raw reads using quality reports from FastQC and removed the specified adapter sequence ATCACCGACTGCCCATAGAGAGGCTGAGAC with Cutadapt (version 1.16). The parameters used for this step were the parameters provided by the submitted of the dataset on Gene Expression Omnibus. The adapter trimmed reads were again assessed for quality by FastQC and were aligned to the reference genome (GRCh37.p13, genecode v19) using a 2-pass method with RNA STAR (Galaxy version 2.7.5b) where other parameters used were default parameters. Following alignment, the generated BAM files were processed using the featureCounts tool (Galaxy version 1.6.4+galaxy2) to get raw counts for each RNA-seq data sample.

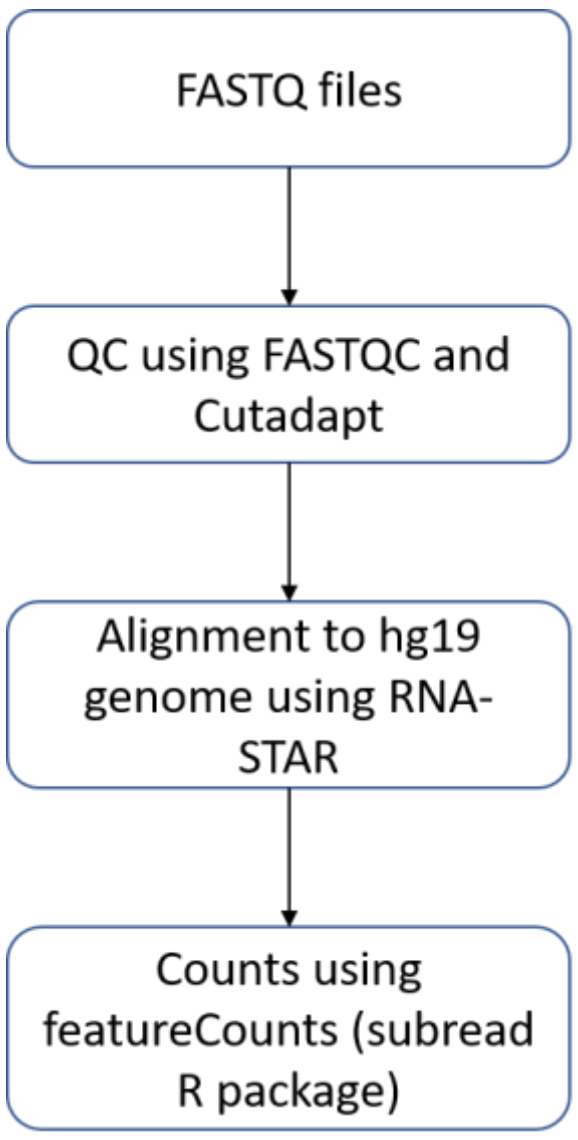

**Figure 1.** Flow of different steps used in processing of the raw RNA-seq fastq files for GSE151352 using Galaxy to get raw counts.

**Supplementary Table 1.** Total number of genes (# genes), number of genes with conflicting direction of expression (# conf. genes), number of differentially expressed genes (# DEGs), number of genes with conflicting direction of expression (# conf. DEGs) and number of true DEGs (# true DEGs) present for all simulation settings averaged over 100 trials.

| Setting ($\sigma$, #studies) | # genes (IN/MIN/FIN) | # conf. genes (IN/MIN/FIN) | # DEGs (IN, MIN, FIN) | # conf. DEGs (IN, MIN, FIN) | # true DEGs |
|---|---|---|---|---|---|
| 0.15, 3 | 16776 ± 19 | 10099 ± 49 | 2268 ± 38, 2611 ± 49, 2441 ± 45 | 0, 243 ± 18, 236 ± 18 | 6436 ± 10 |
| 0.15, 5 | 16871 ± 17 | 13036 ± 43 | 2431 ± 35, 4031 ± 47, 3741 ± 45 | 0, 1365 ± 39, 1337 ± 38 | 6468 ± 9 |
| 0.5, 3 | 17002 ± 45 | 11490 ± 57 | 2098 ± 35, 3879 ± 46, 3741 ± 44 | 0, 1706 ± 35, 1697 ± 35 | 6501 ± 19 |
| 0.5, 5 | 17266 ± 34 | 14824 ± 48 | 1124 ± 31, 4453 ± 48, 4365 ± 46 | 0, 3264 ± 46, 3257 ± 45 | 6585 ± 15 |

**Supplementary Table 2.** Total number of up and down-regulated DEGs identified in per-study differential analysis in 10 random selections of 20 GBM cases and 5 controls from TCGA-GBM dataset.

| TCGA-GBM | DEGs (Up) | DEGs (Down) | Total |
|---|---|---|---|
| Selection 1 | 3112 | 3016 | 6128 |
| Selection 2 | 3323 | 3013 | 6336 |
| Selection 3 | 3204 | 3030 | 6234 |
| Selection 4 | 3285 | 3054 | 6339 |
| Selection 5 | 3267 | 3077 | 6344 |
| Selection 6 | 3249 | 3069 | 6318 |
| Selection 7 | 3293 | 2968 | 6261 |
| Selection 8 | 3792 | 2885 | 6677 |
| Selection 9 | 3311 | 3025 | 6336 |
| Selection 10 | 3229 | 3088 | 6317 |

**Supplementary Table 3.** Number of DEGs identified by the inverse-normal (IN) method in the GBM data application for both scenarios and number of these DEGs present in one, two or all three datasets considered.

| Datasets | DEGs Present in 1 | DEGs Present in 2 | DEGs Present in 3 |
|---|---|---|---|
| GSE123892, GSE151352, TCGA-GBM (all samples) | 1368 | 1085 | 3465 |
| GSE123892, GSE151352, TCGA-GBM (random selection 1) | 921 | 1082 | 3511 |
| GSE123892, GSE151352, TCGA-GBM (random selection 2) | 943 | 1080 | 3477 |
| GSE123892, GSE151352, TCGA-GBM (random selection 3) | 939 | 1079 | 3462 |
| GSE123892, GSE151352, TCGA-GBM (random selection 4) | 965 | 1080 | 3482 |
| GSE123892, GSE151352, TCGA-GBM (random selection 5) | 939 | 1076 | 3540 |
| GSE123892, GSE151352, TCGA-GBM (random selection 6) | 940 | 1100 | 3543 |
| GSE123892, GSE151352, TCGA-GBM (random selection 7) | 972 | 1055 | 3461 |
| GSE123892, GSE151352, TCGA-GBM (random selection 8) | 989 | 1106 | 3513 |
| GSE123892, GSE151352, TCGA-GBM (random selection 9) | 921 | 1083 | 3529 |
| GSE123892, GSE151352, TCGA-GBM (random selection 10) | 995 | 1080 | 3442 |

**Supplementary Table 4.** Number of DEGs identified by the modified inverse-normal (MIN) method in the GBM data application for both scenarios and number of these DEGs present in one, two or all three datasets considered. **a.** For DEGs with the same direction of expression across all three studies. **b.** For DEGs with conflicting direction of expression across studies.

| a. For DEGs with the same direction of expression across all 3 studies | | | |
|---|---|---|---|
| **Datasets** | **DEGs Present in 1** | **DEGs Present in 2** | **DEGs Present in 3** |
| GSE123892, GSE151352, TCGA-GBM (all samples) | 1182 | 1087 | 3623 |
| GSE123892, GSE151352, TCGA-GBM (random selection 1) | 832 | 1081 | 3658 |
| GSE123892, GSE151352, TCGA-GBM (random selection 2) | 844 | 1099 | 3662 |
| GSE123892, GSE151352, TCGA-GBM (random selection 3) | 861 | 1089 | 3616 |
| GSE123892, GSE151352, TCGA-GBM (random selection 4) | 862 | 1088 | 3654 |
| GSE123892, GSE151352, TCGA-GBM (random selection 5) | 845 | 1078 | 3695 |
| GSE123892, GSE151352, TCGA-GBM (random selection 6) | 863 | 1107 | 3689 |
| GSE123892, GSE151352, TCGA-GBM (random selection 7) | 876 | 1062 | 3644 |
| GSE123892, GSE151352, TCGA-GBM (random selection 8) | 887 | 1094 | 3697 |
| GSE123892, GSE151352, TCGA-GBM (random selection 9) | 830 | 1097 | 3702 |
| GSE123892, GSE151352, TCGA-GBM (random selection 10) | 911 | 1090 | 3605 |

| **b. For DEGs with the mismatched direction of expression across 3 studies** | | | |
|---|---|---|---|
| **Datasets** | **DEGs Present in 1** | **DEGs Present in 2** | **DEGs Present in 3** |
| GSE123892, GSE151352, TCGA-GBM (all samples) | 0 | 52 | 181 |
| GSE123892, GSE151352, TCGA-GBM (random selection 1) | 0 | 26 | 140 |
| GSE123892, GSE151352, TCGA-GBM (random selection 2) | 0 | 41 | 171 |
| GSE123892, GSE151352, TCGA-GBM (random selection 3) | 0 | 38 | 149 |
| GSE123892, GSE151352, TCGA-GBM (random selection 4) | 0 | 34 | 166 |
| GSE123892, GSE151352, TCGA-GBM (random selection 5) | 0 | 28 | 154 |
| GSE123892, GSE151352, TCGA-GBM (random selection 6) | 0 | 33 | 140 |
| GSE123892, GSE151352, TCGA-GBM (random selection 7) | 0 | 39 | 173 |
| GSE123892, GSE151352, TCGA-GBM (random selection 8) | 0 | 30 | 179 |
| GSE123892, GSE151352, TCGA-GBM (random selection 9) | 0 | 39 | 165 |
| GSE123892, GSE151352, TCGA-GBM (random selection 10) | 0 | 34 | 164 |

**Supplementary Table 5.** Number of DEGs identified by the fused inverse-normal (FIN) method in the GBM data application for both scenarios and number of these DEGs present in one, two or all three datasets considered. **a.** For DEGs with the same direction of expression across all three studies. **b.** For DEGs with conflicting direction of expression across studies

| **a. For DEGs with the same direction of expression across all 3 studies** | | | |
|---|---|---|---|
| **Datasets** | **DEGs Present in 1** | **DEGs Present in 2** | **DEGs Present in 3** |
| GSE123892, GSE151352, TCGA-GBM (all samples) | 1359 | 1083 | 3461 |
| GSE123892, GSE151352, TCGA-GBM (random selection 1) | 911 | 1080 | 3508 |
| GSE123892, GSE151352, TCGA-GBM (random selection 2) | 933 | 1077 | 3475 |
| GSE123892, GSE151352, TCGA-GBM (random selection 3) | 931 | 1070 | 3457 |
| GSE123892, GSE151352, TCGA-GBM (random selection 4) | 957 | 1077 | 3476 |
| GSE123892, GSE151352, TCGA-GBM (random selection 5) | 931 | 1073 | 3536 |
| GSE123892, GSE151352, TCGA-GBM (random selection 6) | 935 | 1096 | 3540 |
| GSE123892, GSE151352, TCGA-GBM (random selection 7) | 962 | 1051 | 3460 |
| GSE123892, GSE151352, TCGA-GBM (random selection 8) | 978 | 1101 | 3509 |
| GSE123892, GSE151352, TCGA-GBM (random selection 9) | 912 | 1078 | 3526 |
| GSE123892, GSE151352, TCGA-GBM (random selection 10) | 988 | 1076 | 3436 |

| **b. For DEGs with the mismatched direction of expression across 3 studies** | | | |
|---|---|---|---|
| **Datasets** | **DEGs Present in 1** | **DEGs Present in 2** | **DEGs Present in 3** |
| GSE123892, GSE151352, TCGA-GBM (all samples) | 0 | 53 | 182 |
| GSE123892, GSE151352, TCGA-GBM (random selection 1) | 0 | 26 | 140 |
| GSE123892, GSE151352, TCGA-GBM (random selection 2) | 0 | 41 | 171 |
| GSE123892, GSE151352, TCGA-GBM (random selection 3) | 0 | 38 | 149 |
| GSE123892, GSE151352, TCGA-GBM (random selection 4) | 0 | 34 | 166 |
| GSE123892, GSE151352, TCGA-GBM (random selection 5) | 0 | 28 | 154 |
| GSE123892, GSE151352, TCGA-GBM (random selection 6) | 0 | 32 | 140 |
| GSE123892, GSE151352, TCGA-GBM (random selection 7) | 0 | 39 | 172 |
| GSE123892, GSE151352, TCGA-GBM (random selection 8) | 0 | 30 | 179 |
| GSE123892, GSE151352, TCGA-GBM (random selection 9) | 0 | 39 | 165 |
| GSE123892, GSE151352, TCGA-GBM (random selection 10) | 0 | 34 | 164 |

**Supplementary Table 6.** IPA canonical pathways for **a.** up and **b.** down-regulated DEGs present in all three GBM datasets and identified by the FIN method. Ratio denotes the number of DEGs enriched in a pathway to the total number of genes in that pathway.

| Ingenuity Canonical Pathways | -log(BH p-value) | Ratio | z-score |
|---|---|---|---|
| **a. Pathways from up-regulated DEGs** | | | |
| Kinetochore Metaphase Signaling Pathway | 14.1 | 0.376 | 3.413 |
| Hepatic Fibrosis Signaling Pathway | 12.4 | 0.205 | 7.714 |
| Osteoarthritis Pathway | 11.1 | 0.241 | 5.126 |
| Cell Cycle Control of Chromosomal Replication | 8.23 | 0.393 | 4.69 |
| Role of BRCA1 in DNA Damage Response | 7.88 | 0.325 | 2.837 |
| IL-8 Signaling | 7.36 | 0.215 | 5.84 |
| Tumor Microenvironment Pathway | 6.56 | 0.216 | 4.867 |
| Estrogen-mediated S-phase Entry | 6.32 | 0.5 | 2.496 |
| Death Receptor Signaling | 6.19 | 0.272 | 3.4 |
| GP6 Signaling Pathway | 6.17 | 0.244 | 5.385 |
| Acute Phase Response Signaling | 5.98 | 0.206 | 4.596 |
| Colorectal Cancer Metastasis Signaling | 5.9 | 0.182 | 5.376 |
| Dendritic Cell Maturation | 5.76 | 0.201 | 5.831 |
| Glioblastoma Multiforme Signaling | 5.53 | 0.206 | 4.158 |
| ILK Signaling | 5.43 | 0.195 | 4.7 |
| FAT10 Cancer Signaling Pathway | 5.43 | 0.348 | 2 |
| EIF2 Signaling | 5.35 | 0.183 | 2.711 |
| Senescence Pathway | 5.32 | 0.171 | 1.021 |
| Type I Diabetes Mellitus Signaling | 5.27 | 0.234 | 3.441 |
| Regulation Of The Epithelial Mesenchymal Transition By Growth Factors Pathway | 5.16 | 0.191 | 5.667 |
| Production of Nitric Oxide and Reactive Oxygen Species in Macrophages | 5.12 | 0.19 | 5.24 |
| Neuroinflammation Signaling Pathway | 5.03 | 0.163 | 6.112 |
| HOTAIR Regulatory Pathway | 5.01 | 0.2 | 3.772 |
| Cyclins and Cell Cycle Regulation | 5 | 0.259 | 1.528 |
| Signaling by Rho Family GTPases | 4.87 | 0.17 | 5.745 |
| Role of PKR in Interferon Induction and Antiviral Response | 4.84 | 0.22 | 3.411 |
| Leukocyte Extravasation Signaling | 4.58 | 0.181 | 3.889 |
| Pancreatic Adenocarcinoma Signaling | 4.5 | 0.22 | 2.84 |
| Glioma Signaling | 4.44 | 0.218 | 3.317 |
| HIF1α Signaling | 4.43 | 0.176 | 4.333 |
| Tec Kinase Signaling | 4.39 | 0.185 | 4.796 |
| p53 Signaling | 4.28 | 0.224 | 0.535 |
| Inhibition of Angiogenesis by TSP1 | 4.26 | 0.353 | 2.53 |
| Regulation of Cellular Mechanics by Calpain Protease | 3.95 | 0.243 | 1.732 |
| IL-6 Signaling | 3.94 | 0.198 | 4.491 |
| STAT3 Pathway | 3.87 | 0.193 | 2.982 |
| Bladder Cancer Signaling | 3.86 | 0.216 | 2 |
| TNFR1 Signaling | 3.76 | 0.28 | 3.051 |
| Ovarian Cancer Signaling | 3.67 | 0.187 | 3.317 |
| Apoptosis Signaling | 3.67 | 0.21 | 1.789 |
| BEX2 Signaling Pathway | 3.6 | 0.228 | 2.828 |
| Breast Cancer Regulation by Stathmin1 | 3.57 | 0.125 | 6.215 |
| Mitotic Roles of Polo-Like Kinase | 3.54 | 0.242 | 1.069 |
| MSP-RON Signaling In Cancer Cells Pathway | 3.54 | 0.187 | 4.6 |
| Aryl Hydrocarbon Receptor Signaling | 3.51 | 0.182 | 0.894 |
| Systemic Lupus Erythematosus In B Cell Signaling Pathway | 3.47 | 0.149 | 4.841 |
| ATM Signaling | 3.43 | 0.206 | 0.775 |
| TREM1 Signaling | 3.42 | 0.227 | 4.123 |
| Induction of Apoptosis by HIV1 | 3.42 | 0.246 | 2.84 |
| NF-κB Signaling | 3.42 | 0.168 | 4.747 |
| Notch Signaling | 3.25 | 0.297 | 1.414 |
| Myc Mediated Apoptosis Signaling | 3.23 | 0.26 | 2.496 |
| Sphingosine-1-phosphate Signaling | 3.21 | 0.188 | 1.706 |
| Fcγ Receptor-mediated Phagocytosis in Macrophages and Monocytes | 3.18 | 0.202 | 4.359 |
| Glioma Invasiveness Signaling | 3.09 | 0.219 | 2.5 |
| Th2 Pathway | 3.09 | 0.176 | 1.732 |
| Crosstalk between Dendritic Cells and Natural Killer Cells | 3.04 | 0.202 | 3.464 |
| Cytotoxic T Lymphocyte-mediated Apoptosis of Target Cells | 2.94 | 0.294 | 2.449 |
| Regulation Of The Epithelial Mesenchymal Transition In Development Pathway | 2.87 | 0.202 | 3.742 |
| TWEAK Signaling | 2.84 | 0.286 | 1.897 |
| Integrin Signaling | 2.84 | 0.15 | 4.914 |
| Tumoricidal Function of Hepatic Natural Killer Cells | 2.73 | 0.333 | 2.646 |
| Ephrin Receptor Signaling | 2.72 | 0.153 | 3.5 |
| NER Pathway | 2.71 | 0.184 | 2.828 |
| Role of Pattern Recognition Receptors in Recognition of Bacteria and Viruses | 2.71 | 0.162 | 3.742 |
| Wnt/β-catenin Signaling | 2.67 | 0.156 | 1.4 |
| Complement System | 2.67 | 0.27 | 1.414 |
| Protein Kinase A Signaling | 2.52 | 0.125 | 0.686 |
| fMLP Signaling in Neutrophils | 2.5 | 0.172 | 3.207 |
| LPS/IL-1 Mediated Inhibition of RXR Function | 2.47 | 0.142 | 0.943 |
| Actin Cytoskeleton Signaling | 2.41 | 0.141 | 4.2 |
| GM-CSF Signaling | 2.36 | 0.2 | 2.496 |
| VDR/RXR Activation | 2.36 | 0.192 | 0.707 |
| HGF Signaling | 2.34 | 0.167 | 3.357 |
| HMGB1 Signaling | 2.32 | 0.152 | 4.583 |
| Small Cell Lung Cancer Signaling | 2.32 | 0.197 | 2.333 |
| Th1 Pathway | 2.31 | 0.165 | 2.84 |
| Semaphorin Neuronal Repulsive Signaling Pathway | 2.29 | 0.158 | 0.853 |
| PI3K/AKT Signaling | 2.28 | 0.146 | 2.673 |
| Acute Myeloid Leukemia Signaling | 2.23 | 0.18 | 2.53 |
| TNFR2 Signaling | 2.13 | 0.267 | 1.89 |
| Role of NFAT in Regulation of the Immune Response | 2.12 | 0.144 | 4.472 |
| mTOR Signaling | 2.1 | 0.138 | 3.464 |
| Glutathione Redox Reactions I | 2.1 | 0.292 | 2.646 |
| Androgen Signaling | 2.09 | 0.154 | 1.667 |
| Cardiac Hypertrophy Signaling (Enhanced) | 2.09 | 0.115 | 6.934 |
| Toll-like Receptor Signaling | 2.07 | 0.184 | 2.496 |
| iNOS Signaling | 2.07 | 0.222 | 2.828 |
| Mouse Embryonic Stem Cell Pluripotency | 2 | 0.165 | 3.638 |

| **Ingenuity Canonical Pathways** | **-log(BH p-value)** | **Ratio** | **z-score** |
|---|---|---|---|
| **b. Pathways from down-regulated DEGs** | | | |
| Synaptogenesis Signaling Pathway | 24.5 | 0.288 | -8.013 |
| Endocannabinoid Neuronal Synapse Pathway | 16 | 0.359 | -4.727 |
| Role of NFAT in Cardiac Hypertrophy | 14.1 | 0.271 | -6.456 |
| Opioid Signaling Pathway | 13 | 0.247 | -6.04 |
| GNRH Signaling | 10.2 | 0.26 | -5.477 |
| Calcium Signaling | 10.2 | 0.243 | -6.091 |
| G Beta Gamma Signaling | 9.03 | 0.287 | -5.396 |
| Dopamine-DARPP32 Feedback in cAMP Signaling | 8.81 | 0.252 | -4.902 |
| Synaptic Long Term Depression | 8.38 | 0.233 | -5.555 |
| GPCR-Mediated Nutrient Sensing in Enteroendocrine Cells | 8.26 | 0.286 | -5.303 |
| Neuropathic Pain Signaling In Dorsal Horn Neurons | 8.22 | 0.297 | -5.477 |
| Reelin Signaling in Neurons | 7.97 | 0.27 | -5.657 |
| Netrin Signaling | 7.85 | 0.354 | -4.796 |
| Synaptic Long Term Potentiation | 7.38 | 0.256 | -5.209 |
| CCR5 Signaling in Macrophages | 7.09 | 0.287 | -3.464 |
| Melatonin Signaling | 6.93 | 0.319 | -3.545 |
| CXCR4 Signaling | 6.58 | 0.222 | -4.041 |
| Protein Kinase A Signaling | 6.55 | 0.165 | -2.84 |
| Signaling by Rho Family GTPases | 6.44 | 0.19 | -5.24 |
| Corticotropin Releasing Hormone Signaling | 6.17 | 0.228 | -1.095 |
| Cardiac Hypertrophy Signaling | 5.91 | 0.188 | -5.754 |
| Apelin Endothelial Signaling Pathway | 5.85 | 0.243 | -3.657 |
| Sperm Motility | 5.57 | 0.188 | -5.099 |
| Thrombin Signaling | 5.54 | 0.192 | -4.271 |
| Nitric Oxide Signaling in the Cardiovascular System | 5.54 | 0.253 | -4.491 |
| Cardiac β-adrenergic Signaling | 5.53 | 0.22 | -2.837 |
| Androgen Signaling | 5.38 | 0.221 | -4.583 |
| α-Adrenergic Signaling | 5.25 | 0.25 | -3.873 |
| Superpathway of Inositol Phosphate Compounds | 5.24 | 0.191 | -6.083 |
| Type II Diabetes Mellitus Signaling | 4.99 | 0.211 | -3.051 |
| Gαq Signaling | 4.94 | 0.203 | -5.196 |
| Glutamate Receptor Signaling | 4.75 | 0.298 | -1.89 |
| cAMP-mediated signaling | 4.6 | 0.175 | -5.096 |
| Huntington's Disease Signaling | 4.58 | 0.173 | -1.706 |
| nNOS Signaling in Neurons | 4.57 | 0.319 | -3 |
| 3-phosphoinositide Biosynthesis | 4.5 | 0.193 | -5.568 |
| White Adipose Tissue Browning Pathway | 4.44 | 0.209 | -4.426 |
| CREB Signaling in Neurons | 4.42 | 0.134 | -7.778 |
| Insulin Secretion Signaling Pathway | 4.3 | 0.168 | -5.778 |
| FcγRIIB Signaling in B Lymphocytes | 4.27 | 0.253 | -2.449 |
| Renin-Angiotensin Signaling | 4.22 | 0.212 | -4.583 |
| Adrenomedullin signaling pathway | 4.18 | 0.178 | -5.657 |
| D-myo-inositol-5-phosphate Metabolism | 4.18 | 0.191 | -5.385 |
| 14-3-3-mediated Signaling | 4.15 | 0.205 | -3.838 |
| Endothelin-1 Signaling | 3.84 | 0.176 | -4.131 |
| Phospholipase C Signaling | 3.8 | 0.158 | -4.914 |
| Relaxin Signaling | 3.74 | 0.187 | -3.606 |
| D-myo-inositol (1,4,5,6)-Tetrakisphosphate Biosynthesis | 3.74 | 0.19 | -5.099 |
| D-myo-inositol (3,4,5,6)-tetrakisphosphate Biosynthesis | 3.74 | 0.19 | -5.099 |
| Cholecystokinin/Gastrin-mediated Signaling | 3.74 | 0.202 | -4.583 |
| P2Y Purigenic Receptor Signaling Pathway | 3.72 | 0.197 | -4.264 |
| Apelin Cardiomyocyte Signaling Pathway | 3.58 | 0.212 | -4.583 |
| CDK5 Signaling | 3.49 | 0.204 | -2.4 |
| CCR3 Signaling in Eosinophils | 3.46 | 0.194 | -3.742 |
| 3-phosphoinositide Degradation | 3.46 | 0.179 | -5.196 |
| Chemokine Signaling | 3.41 | 0.225 | -3.638 |
| Aldosterone Signaling in Epithelial Cells | 3.38 | 0.177 | -4.359 |
| Cardiac Hypertrophy Signaling (Enhanced) | 3.13 | 0.129 | -7.366 |
| PKCθ Signaling in T Lymphocytes | 3.13 | 0.174 | -3.742 |
| Endocannabinoid Developing Neuron Pathway | 3.11 | 0.191 | -1.606 |
| Sphingosine-1-phosphate Signaling | 3.01 | 0.188 | -2.828 |
| ErbB Signaling | 2.99 | 0.202 | -4.243 |
| UVB-Induced MAPK Signaling | 2.9 | 0.25 | -3.606 |
| p70S6K Signaling | 2.82 | 0.178 | -3.13 |
| Rac Signaling | 2.82 | 0.182 | -4.243 |
| Tec Kinase Signaling | 2.75 | 0.162 | -4.123 |
| RhoA Signaling | 2.73 | 0.179 | -3.13 |
| Pregnenolone Biosynthesis | 2.73 | 0.462 | -2.449 |
| Gαs Signaling | 2.73 | 0.187 | -3.3 |
| IL-1 Signaling | 2.69 | 0.196 | -1.89 |
| fMLP Signaling in Neutrophils | 2.69 | 0.181 | -4.243 |
| Ephrin B Signaling | 2.52 | 0.208 | -3.317 |
| HGF Signaling | 2.51 | 0.175 | -4.243 |
| RhoGDI Signaling | 2.51 | 0.153 | 2.828 |
| Ephrin Receptor Signaling | 2.51 | 0.153 | -4.583 |
| GPCR-Mediated Integration of Enteroendocrine Signaling Exemplified by an L Cell | 2.47 | 0.205 | -0.258 |
| IL-8 Signaling | 2.44 | 0.15 | -4.796 |
| Leptin Signaling in Obesity | 2.41 | 0.203 | -2.236 |
| Histidine Degradation VI | 2.4 | 0.4 | -2.449 |
| B Cell Receptor Signaling | 2.31 | 0.151 | -4.041 |
| Role of NFAT in Regulation of the Immune Response | 2.17 | 0.149 | -3.838 |
| PI3K Signaling in B Lymphocytes | 2.11 | 0.159 | -4.472 |
| Ubiquinol-10 Biosynthesis (Eukaryotic) | 2.09 | 0.353 | -2.449 |
| ERK/MAPK Signaling | 2.09 | 0.144 | -3 |
| Amyotrophic Lateral Sclerosis Signaling | 2.05 | 0.175 | -2.111 |
| RANK Signaling in Osteoclasts | 2.04 | 0.18 | -4 |
| Neuroinflammation Signaling Pathway | 2.03 | 0.13 | -2.785 |
| UVA-Induced MAPK Signaling | 2.01 | 0.173 | -3.464 |

**Supplementary Table 7.** IPA upstream regulator analysis results for **a.** up and **b.** down-regulated DEGs present in all three GBM datasets and identified by the FIN method.

| Upstream Regulator | mean\|logFC\| | Molecule Type | Activation z-score | P-value overlap | BH p-value |
|---|---|---|---|---|---|
| **a.** Upstream regulators for up-regulated DEGs | | | | | |
| TGFB1 | 1.041 | growth factor | 8.971 | 2.51E-88 | 2.08E-84 |
| TP53 | 2.208 | transcription regulator | 3.96 | 9.95E-80 | 4.14E-76 |
| ERBB2 | 1.26 | kinase | 7.269 | 2.60E-66 | 3.60E-63 |
| MYC | 2.687 | transcription regulator | 4.855 | 4.37E-62 | 4.55E-59 |
| CDKN1A | 2.457 | kinase | -4.091 | 2.34E-56 | 1.77E-53 |
| NFKBIA | 1.272 | transcription regulator | 3.344 | 1.80E-43 | 7.47E-41 |
| FOXM1 | 4.83 | transcription regulator | 7.238 | 5.80E-39 | 1.79E-36 |
| EGFR | 4.605 | kinase | 5.001 | 9.51E-37 | 2.64E-34 |
| CCND1 | 1.251 | transcription regulator | 4.578 | 2.88E-33 | 5.44E-31 |
| YAP1 | 1.162 | transcription regulator | 5.297 | 7.46E-32 | 1.29E-29 |
| TWIST1 | 1.889 | transcription regulator | 4.192 | 7.09E-30 | 1.02E-27 |
| E2F1 | 1.146 | transcription regulator | 5.853 | 1.04E-28 | 1.35E-26 |
| TBX2 | 3.103 | transcription regulator | 4.433 | 7.35E-28 | 8.86E-26 |
| NUPR1 | 1.918 | transcription regulator | -3.927 | 1.43E-26 | 1.51E-24 |
| AR | 2.049 | ligand-dependent nuclear receptor | 3.669 | 1.19E-25 | 1.15E-23 |
| JUN | 1.495 | transcription regulator | 5.72 | 2.87E-25 | 2.57E-23 |
| NR1H3 | 1.074 | ligand-dependent nuclear receptor | 2.203 | 1.23E-24 | 1.05E-22 |
| AKT1 | 1.066 | kinase | 4.256 | 3.66E-24 | 2.92E-22 |
| VEGFA | 3.776 | growth factor | 6.513 | 1.21E-23 | 9.28E-22 |
| SPI1 | 1.254 | transcription regulator | 1.876 | 1.63E-23 | 1.22E-21 |
| TCF3 | 1.887 | transcription regulator | -0.806 | 9.61E-23 | 6.50E-21 |
| FAS | 1.754 | transmembrane receptor | -2.084 | 2.13E-22 | 1.42E-20 |
| TGFB2 | 1.198 | growth factor | 3.424 | 5.11E-22 | 3.32E-20 |
| CD44 | 2.568 | other | 4.541 | 2.28E-20 | 1.27E-18 |
| APOE | 1.358 | transporter | -4.791 | 2.73E-19 | 1.39E-17 |
| IRF1 | 2.009 | transcription regulator | 5.486 | 9.97E-19 | 4.74E-17 |
| CEBPB | 1.262 | transcription regulator | 5.281 | 1.01E-18 | 4.77E-17 |
| FN1 | 2.971 | enzyme | 5.104 | 2.56E-17 | 1.10E-15 |
| CDK4 | 3.341 | kinase | 0.469 | 6.10E-17 | 2.51E-15 |
| ANGPT2 | 3.221 | growth factor | 6.136 | 6.54E-17 | 2.68E-15 |
| CCN2 | 1.487 | growth factor | 4.292 | 5.92E-16 | 2.16E-14 |
| E2F2 | 5.191 | transcription regulator | 2.494 | 8.16E-16 | 2.95E-14 |
| EZH2 | 3.461 | transcription regulator | 0.551 | 8.72E-16 | 3.14E-14 |
| NOTCH3 | 2.12 | transcription regulator | 4.045 | 1.01E-15 | 3.60E-14 |
| IRF7 | 1.324 | transcription regulator | 6.854 | 1.06E-15 | 3.73E-14 |
| RBL1 | 1.244 | transcription regulator | -4.344 | 4.95E-15 | 1.62E-13 |
| F2R | 3.98 | G-protein coupled receptor | 5.338 | 6.13E-15 | 2.00E-13 |
| BIRC5 | 4.835 | other | -1.709 | 1.25E-14 | 3.99E-13 |
| S100A9 | 3.908 | other | 3.047 | 1.42E-14 | 4.51E-13 |
| CD40 | 1.101 | transmembrane receptor | 4.744 | 1.43E-14 | 4.51E-13 |
| NAMPT | 2.502 | cytokine | 2.682 | 1.61E-14 | 5.07E-13 |
| MYBL2 | 6.787 | transcription regulator | 2.79 | 2.89E-14 | 8.80E-13 |
| ABCB4 | 2.036 | transporter | -4.078 | 4.01E-14 | 1.18E-12 |
| JAG1 | 2.726 | growth factor | 4.426 | 7.45E-14 | 2.14E-12 |
| COL18A1 | 1.65 | other | -4.828 | 7.84E-14 | 2.22E-12 |
| PLK1 | 2.524 | kinase | -2.02 | 8.98E-14 | 2.51E-12 |
| ACVRL1 | 1.465 | kinase | -0.024 | 1.80E-13 | 4.82E-12 |
| S100A6 | 2.611 | transporter | 0.943 | 2.05E-13 | 5.45E-12 |
| CIP2A | 1.493 | other | -1.763 | 2.76E-13 | 7.26E-12 |
| HOXA10 | 7.693 | transcription regulator | -2.33 | 2.84E-13 | 7.44E-12 |
| RARA | 1.599 | ligand-dependent nuclear receptor | 3.454 | 3.26E-13 | 8.42E-12 |
| TEAD4 | 2.789 | transcription regulator | 4.204 | 3.84E-13 | 9.81E-12 |
| IL33 | 1.493 | cytokine | 5.613 | 4.17E-13 | 1.06E-11 |
| DCN | 1.381 | other | -1.12 | 7.13E-13 | 1.77E-11 |
| S100A8 | 4.872 | other | 2.378 | 7.63E-13 | 1.89E-11 |
| ID3 | 2.405 | transcription regulator | -0.307 | 1.94E-12 | 4.46E-11 |
| BCL6 | 1.086 | transcription regulator | -1.729 | 4.74E-12 | 1.04E-10 |
| CCN1 | 2.124 | other | 3.663 | 7.29E-12 | 1.56E-10 |
| MYD88 | 1.788 | other | 6.715 | 1.37E-11 | 2.83E-10 |
| TREM1 | 4.669 | transmembrane receptor | 3.236 | 2.24E-11 | 4.50E-10 |
| TNFSF13B | 1.681 | cytokine | 3.489 | 2.32E-11 | 4.64E-10 |
| ENG | 1.746 | transmembrane receptor | 0.342 | 2.82E-11 | 5.59E-10 |
| HMOX1 | 2.464 | enzyme | -1.425 | 4.55E-11 | 8.73E-10 |
| TEAD2 | 3.281 | transcription regulator | 4 | 7.47E-11 | 1.40E-09 |
| SPP1 | 1.055 | cytokine | 5.002 | 9.83E-11 | 1.79E-09 |
| IL10RA | 1.235 | transmembrane receptor | -0.431 | 1.01E-10 | 1.84E-09 |
| SMAD1 | 1.086 | transcription regulator | 3.773 | 1.03E-10 | 1.87E-09 |
| CXCR4 | 2.307 | G-protein coupled receptor | 2.791 | 1.18E-10 | 2.12E-09 |
| ETS1 | 1.474 | transcription regulator | 4.918 | 1.43E-10 | 2.55E-09 |
| NOX4 | 2.202 | enzyme | 3.208 | 2.56E-10 | 4.46E-09 |
| MAP3K1 | 1.417 | kinase | 4.376 | 2.69E-10 | 4.66E-09 |
| ATF3 | 1.086 | transcription regulator | -1.834 | 3.38E-10 | 5.79E-09 |
| THBS1 | 2.392 | other | 0.938 | 4.76E-10 | 8.00E-09 |
| WWTR1 | 2.559 | transcription regulator | 2.275 | 5.97E-10 | 9.83E-09 |
| GLIS2 | 1.085 | transcription regulator | -3.45 | 6.95E-10 | 1.14E-08 |
| RUNX3 | 1.993 | transcription regulator | -3.767 | 8.63E-10 | 1.39E-08 |
| KLF6 | 1.287 | transcription regulator | 1.665 | 9.32E-10 | 1.49E-08 |
| TFAP2A | 3.639 | transcription regulator | -1.105 | 9.69E-10 | 1.54E-08 |
| PLAU | 4.022 | peptidase | 2.262 | 1.38E-09 | 2.14E-08 |
| CHEK1 | 2.576 | kinase | 1.233 | 1.75E-09 | 2.67E-08 |
| PTGER4 | 1.554 | G-protein coupled receptor | -1.729 | 2.73E-09 | 4.06E-08 |
| ETV4 | 3.275 | transcription regulator | 3.774 | 3.55E-09 | 5.16E-08 |
| S100A4 | 3.034 | other | 2.341 | 4.52E-09 | 6.43E-08 |
| MMP9 | 6.815 | peptidase | 3.108 | 5.04E-09 | 7.08E-08 |
| TLR3 | 1.497 | transmembrane receptor | 5.937 | 5.38E-09 | 7.53E-08 |
| TEAD3 | 2.377 | transcription regulator | 3.771 | 6.51E-09 | 9.03E-08 |
| SOX2 | 1.572 | transcription regulator | 0.781 | 7.82E-09 | 1.07E-07 |
| TNFRSF1A | 1.995 | transmembrane receptor | 1.079 | 1.15E-08 | 1.52E-07 |
| ANXA2 | 3.997 | other | -1.574 | 1.25E-08 | 1.64E-07 |
| SNAI2 | 2.56 | transcription regulator | 2.345 | 1.34E-08 | 1.75E-07 |
| HMGA1 | 1.256 | transcription regulator | 1.163 | 1.44E-08 | 1.86E-07 |
| ITGA5 | 2.844 | transmembrane receptor | 2.102 | 1.47E-08 | 1.90E-07 |
| PDGFC | 1.867 | growth factor | 3.388 | 1.79E-08 | 2.30E-07 |
| TYROBP | 1.505 | transmembrane receptor | 2.2 | 2.24E-08 | 2.83E-07 |
| DLL4 | 1.581 | other | -0.357 | 2.27E-08 | 2.86E-07 |
| PRDM1 | 1.411 | transcription regulator | -1.626 | 2.51E-08 | 3.15E-07 |
| SPARC | 2.132 | other | -0.419 | 2.68E-08 | 3.36E-07 |
| ZEB1 | 1.431 | transcription regulator | 1.684 | 4.32E-08 | 5.24E-07 |
| NCF1 | 1.26 | enzyme | 2.905 | 5.57E-08 | 6.62E-07 |
| IGFBP2 | 4.882 | other | 2.823 | 6.06E-08 | 7.13E-07 |
| SOX4 | 2.495 | transcription regulator | 3.934 | 6.18E-08 | 7.26E-07 |
| TIMP1 | 3.632 | cytokine | 1.604 | 1.11E-07 | 1.24E-06 |
| E2F5 | 1.566 | transcription regulator | 2 | 1.15E-07 | 1.28E-06 |
| TLR2 | 1.287 | transmembrane receptor | 1.861 | 1.60E-07 | 1.74E-06 |
| CXCL8 | 3.465 | cytokine | 2.976 | 2.01E-07 | 2.15E-06 |
| ASCL1 | 1.684 | transcription regulator | 2.383 | 2.03E-07 | 2.16E-06 |
| PARP9 | 1.68 | enzyme | 2.373 | 2.29E-07 | 2.42E-06 |
| NEDD9 | 1.814 | other | 1.973 | 2.29E-07 | 2.42E-06 |
| ZMYND10 | 1.329 | other | -0.342 | 2.45E-07 | 2.55E-06 |
| ACKR3 | 1.907 | G-protein coupled receptor | 2.027 | 2.45E-07 | 2.55E-06 |
| MSTN | 1.995 | growth factor | 0.366 | 2.90E-07 | 2.98E-06 |
| SPHK1 | 1.69 | kinase | 1.646 | 3.16E-07 | 3.22E-06 |
| GMNN | 1.505 | transcription regulator | -3.748 | 3.20E-07 | 3.26E-06 |
| WNT5A | 1.846 | cytokine | 1.597 | 3.22E-07 | 3.27E-06 |
| CTSB | 1.196 | peptidase | 2.408 | 3.31E-07 | 3.35E-06 |
| LGALS3 | 3.025 | other | 0.747 | 3.35E-07 | 3.39E-06 |
| BAX | 1.466 | transporter | 2.256 | 3.61E-07 | 3.62E-06 |
| POSTN | 6.219 | other | 2.589 | 3.61E-07 | 3.62E-06 |
| DDB2 | 1.301 | other | 0.354 | 5.65E-07 | 5.46E-06 |
| IFIH1 | 1.048 | enzyme | 3.142 | 6.04E-07 | 5.77E-06 |
| CDK2 | 2.938 | kinase | 2.496 | 7.22E-07 | 6.82E-06 |
| ACTL6A | 1.54 | other | 0.905 | 8.51E-07 | 7.96E-06 |
| MAP3K14 | 1.231 | kinase | 2.881 | 8.79E-07 | 8.19E-06 |
| NFKB2 | 1.33 | transcription regulator | 3.218 | 1.09E-06 | 1.01E-05 |
| RUNX1 | 1.627 | transcription regulator | 1.428 | 1.10E-06 | 1.01E-05 |
| OSMR | 2.989 | transmembrane receptor | 1.4 | 1.13E-06 | 1.04E-05 |
| CASP3 | 1.189 | peptidase | 1.249 | 1.32E-06 | 1.20E-05 |
| GDF15 | 5.161 | growth factor | 0.765 | 1.48E-06 | 1.33E-05 |
| PGF | 2.975 | growth factor | 2.245 | 1.66E-06 | 1.46E-05 |
| PLAUR | 2.196 | transmembrane receptor | 2.61 | 1.66E-06 | 1.46E-05 |
| MUC1 | 2.76 | other | 2.782 | 1.87E-06 | 1.63E-05 |
| BCL6B | 1.668 | transcription regulator | 1.309 | 1.87E-06 | 1.63E-05 |
| ADAMTS12 | 2.3 | peptidase | -3.138 | 1.88E-06 | 1.63E-05 |
| PTGS1 | 2.051 | enzyme | 1.5 | 2.30E-06 | 1.97E-05 |
| TRAF3IP2 | 1.532 | other | 3.939 | 2.71E-06 | 2.29E-05 |
| HSPB1 | 1.863 | other | 0.921 | 2.95E-06 | 2.47E-05 |
| PDGFRA | 2.422 | kinase | 2.21 | 3.02E-06 | 2.51E-05 |
| TYMS | 1.54 | enzyme | -2.213 | 3.10E-06 | 2.55E-05 |
| CEBPD | 2.348 | transcription regulator | 0.231 | 3.11E-06 | 2.56E-05 |
| ICAM1 | 2.224 | transmembrane receptor | 3.418 | 3.40E-06 | 2.79E-05 |
| MEX3A | 3.168 | other | 1.387 | 3.80E-06 | 3.09E-05 |
| MMP2 | 3.159 | peptidase | 2.053 | 3.80E-06 | 3.09E-05 |
| ADAM12 | 2.471 | peptidase | -0.269 | 4.19E-06 | 3.40E-05 |
| ANXA1 | 3.445 | enzyme | -1.1 | 4.31E-06 | 3.48E-05 |
| HLX | 1.738 | transcription regulator | -1.66 | 4.31E-06 | 3.48E-05 |
| NT5E | 1.606 | phosphatase | -1.134 | 5.34E-06 | 4.24E-05 |
| GPX1 | 1.407 | enzyme | -3.532 | 5.80E-06 | 4.59E-05 |
| IGFBP7 | 2.448 | transporter | -0.174 | 5.91E-06 | 4.64E-05 |
| EDNRA | 2.101 | transmembrane receptor | 2.828 | 5.91E-06 | 4.64E-05 |
| ETV1 | 2.044 | transcription regulator | 2.158 | 6.60E-06 | 5.13E-05 |
| TGFB1I1 | 3.679 | transcription regulator | 0.452 | 6.60E-06 | 5.13E-05 |
| SAMSN1 | 1.112 | other | 4.583 | 6.61E-06 | 5.13E-05 |
| CIITA | 1.126 | transcription regulator | 1.576 | 7.27E-06 | 5.59E-05 |
| LATS2 | 1.081 | kinase | 0.971 | 7.27E-06 | 5.59E-05 |
| PIM1 | 1.032 | kinase | 2.179 | 7.28E-06 | 5.60E-05 |
| AURKB | 5.485 | kinase | -1.177 | 9.02E-06 | 6.78E-05 |
| IFI16 | 1.477 | transcription regulator | 3.262 | 9.50E-06 | 7.12E-05 |
| KDR | 1.413 | kinase | 1.105 | 1.02E-05 | 7.60E-05 |
| SHC1 | 1.751 | other | -0.967 | 1.60E-05 | 0.000115 |
| PSMB9 | 1.808 | peptidase | 0.391 | 1.67E-05 | 0.000118 |
| SERPINH1 | 3.581 | other | 2.401 | 1.67E-05 | 0.000118 |
| CPXM1 | 4.813 | peptidase | 2.236 | 1.74E-05 | 0.000121 |
| LEF1 | 1.707 | transcription regulator | 1.929 | 1.90E-05 | 0.000132 |
| ZNF217 | 1.274 | transcription regulator | -1.846 | 2.03E-05 | 0.00014 |
| AIF1 | 1.007 | other | 2.607 | 2.16E-05 | 0.000149 |
| SPRY2 | 1.039 | other | -0.992 | 2.27E-05 | 0.000155 |
| ARHGAP31 | 1.041 | other | -0.905 | 2.81E-05 | 0.000189 |
| SOCS3 | 2.869 | phosphatase | -1.99 | 3.27E-05 | 0.000217 |
| TRIB3 | 2.785 | kinase | -1.897 | 3.29E-05 | 0.000218 |
| TNFAIP3 | 1.782 | enzyme | -2.028 | 3.63E-05 | 0.000237 |
| PLAT | 2.471 | peptidase | 2.921 | 3.86E-05 | 0.000251 |
| AEBP1 | 2.812 | peptidase | -0.403 | 3.89E-05 | 0.000252 |
| CYBA | 1.593 | enzyme | 2.414 | 3.89E-05 | 0.000252 |
| RUVBL1 | 1.297 | transcription regulator | 1.808 | 3.92E-05 | 0.000252 |
| SNHG1 | 1.044 | other | -0.243 | 3.92E-05 | 0.000252 |
| LAMC1 | 2.891 | other | -2.616 | 4.03E-05 | 0.000256 |
| ID4 | 1.659 | transcription regulator | -1.951 | 4.03E-05 | 0.000256 |
| HOXA7 | 7.114 | transcription regulator | 2.547 | 4.67E-05 | 0.000294 |
| C3AR1 | 1.143 | G-protein coupled receptor | 2.107 | 5.03E-05 | 0.000317 |
| EPHA2 | 2.476 | kinase | 1.977 | 5.20E-05 | 0.000326 |
| NGFR | 2.284 | transmembrane receptor | -0.328 | 5.22E-05 | 0.000326 |
| TNFRSF1B | 1.187 | transmembrane receptor | 3.29 | 5.64E-05 | 0.000348 |
| MCM7 | 1.516 | enzyme | NA | 5.68E-05 | 0.000348 |
| CKS1B | 1.363 | kinase | 2.215 | 5.68E-05 | 0.000348 |
| C3 | 1.272 | peptidase | 2.563 | 6.00E-05 | 0.000367 |
| C1QA | 2.385 | other | 1.237 | 6.51E-05 | 0.000395 |
| CRNDE | 2.876 | other | 2.37 | 6.99E-05 | 0.000424 |
| HAS2 | 3.55 | enzyme | 0.808 | 7.98E-05 | 0.000475 |
| CDK1 | 3.762 | kinase | 2.213 | 8.07E-05 | 0.000478 |
| RND3 | 2.011 | enzyme | 1.945 | 8.07E-05 | 0.000478 |
| LTBR | 1.001 | transmembrane receptor | 1.508 | 8.09E-05 | 0.000479 |
| MAPK7 | 1.248 | kinase | 1.631 | 8.83E-05 | 0.00052 |
| CLU | 1.038 | other | 0.913 | 8.92E-05 | 0.000523 |
| HOXD10 | 8.605 | transcription regulator | 1.343 | 9.14E-05 | 0.000534 |
| SOX11 | 3.377 | transcription regulator | 1.646 | 9.89E-05 | 0.000575 |
| INHBB | 1.78 | growth factor | 2.8 | 0.000108 | 0.000628 |
| EFNA2 | 1.755 | kinase | -2.673 | 0.000109 | 0.000628 |
| ID1 | 1.214 | transcription regulator | 2.043 | 0.000109 | 0.00063 |
| MAFB | 1.242 | transcription regulator | 1.323 | 0.000115 | 0.000665 |
| COL1A1 | 4.721 | other | 1.934 | 0.000117 | 0.000667 |
| GAS2L3 | 3.14 | other | -2.63 | 0.000117 | 0.000667 |
| CBX3 | 1.288 | transcription regulator | NA | 0.000121 | 0.000689 |
| PRKD1 | 1.199 | kinase | 3.532 | 0.000126 | 0.000714 |
| SMO | 2.739 | G-protein coupled receptor | NA | 0.000135 | 0.00076 |
| DES | 2.527 | other | -1.98 | 0.000142 | 0.000789 |
| ELN | 2.609 | other | 0.896 | 0.000149 | 0.000822 |
| CTSS | 1.594 | peptidase | 1.026 | 0.000151 | 0.000834 |
| MEOX2 | 4.66 | transcription regulator | -2.474 | 0.000156 | 0.000859 |
| RIPK1 | 1.161 | kinase | 0.632 | 0.00018 | 0.000978 |
| MELK | 5.434 | kinase | 1.896 | 0.000184 | 0.00098 |
| FZD7 | 2.772 | G-protein coupled receptor | 1 | 0.000184 | 0.00098 |
| BRCA2 | 2.764 | transcription regulator | -1.154 | 0.000184 | 0.00098 |
| HELLS | 1.894 | enzyme | 2 | 0.000185 | 0.00098 |
| NEDD4 | 1.549 | enzyme | -1.671 | 0.000185 | 0.00098 |
| PTTG1 | 3.247 | transcription regulator | -0.238 | 0.000215 | 0.00113 |
| STK40 | 1.112 | kinase | 1.589 | 0.000253 | 0.00131 |
| GAS5 | 1.03 | other | -1.604 | 0.000253 | 0.00131 |
| LOX | 4.548 | enzyme | 1.906 | 0.000253 | 0.00131 |
| PDGFRB | 1.307 | kinase | 0.842 | 0.000258 | 0.00133 |
| ITGB2 | 1.448 | transmembrane receptor | 0.971 | 0.000278 | 0.00143 |
| EFNA4 | 1.655 | kinase | -3.317 | 0.00032 | 0.00161 |
| HEY1 | 1.792 | transcription regulator | -0.97 | 0.000354 | 0.00176 |
| GAPDH | 1.067 | enzyme | -2.109 | 0.000355 | 0.00177 |
| DPP4 | 2.169 | peptidase | 0.283 | 0.000399 | 0.00198 |
| SULF2 | 1.281 | enzyme | 1.391 | 0.000415 | 0.00204 |
| ACTB | 1.024 | other | 0.956 | 0.000415 | 0.00204 |
| BIRC3 | 1.523 | enzyme | -0.216 | 0.000484 | 0.00235 |
| USP18 | 1.777 | peptidase | -2.959 | 0.000487 | 0.00235 |
| MECOM | 1.255 | transcription regulator | 0.577 | 0.000487 | 0.00235 |
| CD14 | 2.631 | transmembrane receptor | 3.24 | 0.000497 | 0.00236 |
| TIMP3 | 1.46 | other | -2.354 | 0.000507 | 0.0024 |
| MKI67 | 5.201 | other | 1.216 | 0.000516 | 0.00242 |
| PARP14 | 1.209 | enzyme | 1.387 | 0.000556 | 0.00257 |
| ITGA4 | 2.774 | transmembrane receptor | -0.129 | 0.000556 | 0.00257 |
| ATF5 | 1.443 | transcription regulator | 1.969 | 0.000556 | 0.00257 |
| CISH | 2.232 | other | -1.175 | 0.000598 | 0.00272 |
| CGAS | 1.247 | enzyme | 2.815 | 0.000651 | 0.00294 |
| CDH2 | 1.052 | other | 0.577 | 0.000655 | 0.00294 |
| NRP1 | 2.015 | transmembrane receptor | 0.128 | 0.000655 | 0.00294 |
| ADM | 3.417 | other | -2.099 | 0.000735 | 0.00328 |
| MMP14 | 3.434 | peptidase | -1.195 | 0.000807 | 0.00359 |
| NLRC5 | 1.94 | transcription regulator | 1.926 | 0.000832 | 0.00368 |
| PROCR | 1.694 | other | 0.788 | 0.000862 | 0.0038 |
| BGN | 1.918 | other | 0.239 | 0.000862 | 0.0038 |
| CALR | 1.239 | transcription regulator | 1.525 | 0.000902 | 0.00396 |
| CD36 | 1.975 | transmembrane receptor | 4.202 | 0.000908 | 0.00398 |
| MCL1 | 1.314 | transporter | 0.871 | 0.000954 | 0.00414 |
| H1-2 | 2.235 | other | -0.42 | 0.000978 | 0.00423 |
| DNASE2 | 1.292 | enzyme | -2.784 | 0.00102 | 0.0044 |
| SOX9 | 1.516 | transcription regulator | 2.165 | 0.00103 | 0.00444 |
| CCR1 | 1.03 | G-protein coupled receptor | 0.895 | 0.00112 | 0.00476 |
| TMPO | 1.043 | other | 2 | 0.00113 | 0.00476 |
| LGALS1 | 2.065 | other | 2.275 | 0.00138 | 0.00574 |
| TNFAIP6 | 1.941 | other | 1.915 | 0.00152 | 0.00624 |
| HNRNPAB | 1.296 | enzyme | 2 | 0.00153 | 0.00624 |
| YBX1 | 1.7 | transcription regulator | 1.719 | 0.00169 | 0.00687 |
| EBF1 | 1.316 | transcription regulator | 3.057 | 0.00197 | 0.00772 |
| AURKA | 2.84 | kinase | 1.023 | 0.00198 | 0.00772 |
| IGFBP5 | 2.229 | other | 2.007 | 0.00201 | 0.00778 |
| ACE | 2.002 | peptidase | 0.816 | 0.00201 | 0.00778 |
| HOXA5 | 7.06 | transcription regulator | 2.361 | 0.00201 | 0.00778 |
| HAND2 | 5.899 | transcription regulator | -0.444 | 0.00205 | 0.00795 |
| EFEMP1 | 2.47 | enzyme | -1.982 | 0.00212 | 0.00807 |
| TSPO | 1.285 | transmembrane receptor | -1 | 0.00212 | 0.00807 |
| PLA2G5 | 2.165 | enzyme | 1.98 | 0.00212 | 0.00807 |
| NR5A2 | 4.148 | ligand-dependent nuclear receptor | 0.978 | 0.00246 | 0.00929 |
| SOX6 | 1.243 | transcription regulator | 0.862 | 0.0026 | 0.00976 |

| Upstream Regulator | mean\|logFC\| | Molecule Type | Activation z-score | P-value overlap | BH p-value |
|---|---|---|---|---|---|
| **b.** Upstream regulators for down-regulated DEGs | | | | | |
| TCF7L2 | NA | transcription regulator | -9.736 | 9.79E-31 | 4.22E-27 |
| MAPT | -1.398 | other | 0.728 | 2.66E-24 | 5.74E-21 |
| levodopa | NA | chemical - endogenous mammalian | -3.748 | 2.00E-19 | 2.88E-16 |
| HTT | NA | transcription regulator | 1.919 | 7.36E-17 | 6.91E-14 |
| FMR1 | NA | translation regulator | 5.211 | 8.02E-17 | 6.91E-14 |
| SNCA | -4.209 | enzyme | -1.176 | 3.76E-15 | 2.52E-12 |
| REST | NA | transcription regulator | 4.284 | 4.09E-15 | 2.52E-12 |
| BDNF | NA | growth factor | -5.649 | 6.51E-12 | 3.51E-09 |
| DSCAML1 | -2.164 | other | -3.651 | 9.02E-12 | 4.32E-09 |
| HDAC4 | NA | transcription regulator | -0.863 | 1.16E-10 | 4.98E-08 |
| DSCAM | NA | other | -4.163 | 2.22E-10 | 8.70E-08 |
| MKNK1 | NA | kinase | -5.477 | 3.02E-10 | 1.08E-07 |
| topotecan | NA | chemical drug | 4.753 | 4.85E-10 | 1.50E-07 |
| CREB1 | NA | transcription regulator | 0.459 | 4.86E-10 | 1.50E-07 |
| PHF21A | NA | other | -2.51 | 1.43E-09 | 4.12E-07 |
| MECP2 | NA | transcription regulator | -2.411 | 6.96E-09 | 1.76E-06 |
| tetrodotoxin | NA | chemical drug | 2.219 | 1.34E-08 | 3.21E-06 |
| Calmodulin | NA | group | -2.059 | 1.87E-08 | 4.25E-06 |
| PSEN1 | -1.141 | peptidase | -0.722 | 4.16E-08 | 8.96E-06 |
| GRIN3A | NA | ion channel | 5.292 | 8.98E-08 | 1.84E-05 |
| DMD | NA | other | -1.32 | 3.12E-07 | 6.11E-05 |
| APP | -1.079 | other | 0.476 | 1.92E-06 | 0.000359 |
| NFASC | -2.903 | other | -2.952 | 3.81E-06 | 0.000684 |
| tazemetostat | NA | chemical drug | -2.929 | 4.89E-06 | 0.000842 |
| SLC30A3 | NA | transporter | -2.433 | 1.96E-05 | 0.00325 |
| SP2509 | NA | chemical reagent | -2.771 | 2.79E-05 | 0.00444 |
| MFSD2A | NA | transporter | 2.949 | 2.89E-05 | 0.00444 |
| JAK1/2 | NA | group | -4.123 | 4.19E-05 | 0.0061 |
| D-2-amino-5-phosphonovaleric acid | NA | chemical reagent | 0.152 | 4.39E-05 | 0.0061 |

**Supplementary Table 8.** Number of common DEGs identified by the FIN method in both scenarios considered for TCGA-GBM data. On average about 94% of the DEGs obtained when randomly selected subset was considered were also found in DEGs identified using the full TCGA-GBM dataset.

| Scenario 2 | Common DEGs | Total # DEGs | %common |
|---|---|---|---|
| Selection 1 | 5339 | 5665 | 94.25 |
| Selection 2 | 5406 | 5697 | 94.89 |
| Selection 3 | 5308 | 5645 | 94.03 |
| Selection 4 | 5382 | 5710 | 94.26 |
| Selection 5 | 5358 | 5722 | 93.64 |
| Selection 6 | 5336 | 5743 | 92.91 |
| Selection 7 | 5382 | 5684 | 94.69 |
| Selection 8 | 5400 | 5797 | 93.15 |
| Selection 9 | 5352 | 5720 | 93.57 |
| Selection 10 | 5415 | 5698 | 95.03 |